\documentclass[preprint2,twocolumn]{aastex61}
\usepackage{booktabs}
\usepackage{amsmath}
\usepackage{array}

\newcommand{\msun}{M$_\odot$}
\newcommand{\lsun}{L$_\odot$}
\newcommand{\mmsun}{\rm M_\odot}
\newcommand{\mlsun}{\rm L_\odot}
\newcommand{\mbh}{M$_\bullet$}
\newcommand{\mmbh}{\rm M_\bullet}
\newcommand{\etal}{\mbox{\it{et al.}~}}
\newcommand{\Htwo}{H$_{2}$}
\newcommand{\HI}{\ion{H}{1}}
\newcommand{\HII}{\ion{H}{2}}
\newcommand{\HeI}{\ion{He}{1}}
\newcommand{\HeII}{\ion{He}{2}}
\newcommand{\Fe}{[\ion{Fe}{2}]}
\newcommand{\SII}{[\ion{S}{2}]}
\newcommand{\PII}{[\ion{P}{2}]}
\newcommand{\SIII}{[\ion{S}{3}]}
\newcommand{\OI}{[\ion{O}{1}]}
\newcommand{\OII}{[\ion{O}{2}]}
\newcommand{\OIII}{[\ion{O}{3}]}
\newcommand{\NII}{[\ion{N}{2}]}
\newcommand{\Ha}{H$\alpha${}}
\newcommand{\Hb}{H$\beta${}}
\newcommand{\Hg}{H$\gamma${}}
\newcommand{\Fel}{[\ion{Fe}{2}]~$\lambda$}

\newcommand{\pwr}[1]{$^{#1}$}
\newcommand{\pab}{Pa$\beta${}}
\newcommand{\pag}{Pa$\gamma${}}
\newcommand{\brg}{Br$\gamma${}}

\newcommand{\ecs}{erg~cm\pwr{-2}~s\pwr{-1}}
\newcommand{\ecsf}{$\times$ 10\pwr{-16}~erg~cm\pwr{-2}~s\pwr{-1}}
\newcommand{\kms}{km~s\pwr{-1}}
\newcommand{\um}{$\mu$m}
\makeatletter
\newcommand*{\rom}[1]{\expandafter\@slowromancap\romannumeral #1@}
\makeatother
\usepackage{natbib}

\shorttitle{IC~630 Nuclear Gas}
\shortauthors{Durr\'{e} \etal}
\accepted{10 March 2017}

\begin{document}

\title{IC~630: Piercing the Veil of the Nuclear Gas}
%
\author[0000-0002-2126-3905]{Mark Durr\'{e}}
\author[0000-0003-3820-1740]{Jeremy Mould}
\author[0000-0003-1318-8631]{Marc Schartmann}
\affil{Centre for Astrophysics and Supercomputing, Swinburne University of Technology, P.O. Box 218, Hawthorn, Victoria 3122, Australia}
\email{2064944@swin.edu.au}
\author[0000-0002-9413-4186]{Syed Ashraf Uddin}
\affil{Purple Mountain Observatory, Chinese Academy of Sciences, Nanjing, Jiangshu 210008, China}
\author[0000-0002-9975-1829]{Garrett Cotter}
\affil{Oxford Astrophysics, Denys Wilkinson Building, Keble Road, Oxford, OX1 3RH, United Kingdom}
\begin{abstract}
IC~630 is a nearby early-type galaxy with a mass of $6 \times 10^{10}$ \msun with an intense burst of recent (6 Myr) star formation. It shows strong nebular emission lines, with radio and X-ray emission, which classifies it as an AGN. With VLT-SINFONI and Gemini North-NIFS adaptive optics observations (plus supplementary ANU 2.3m WiFeS optical IFU observations), the excitation diagnostics of the nebular emission species show no sign of standard AGN engine excitation; the stellar velocity dispersion also indicate that a super-massive black hole (if one is present) is small ($\mmbh = 2.25 \times 10^{5}~\mmsun$). The luminosity at all wavelengths is consistent with star formation at a rate of about 1--2 \msun/yr. We measure gas outflows driven by star formation at a rate of 0.18 \msun/yr in a face-on truncated cone geometry. We also observe a nuclear cluster or disk and other clusters. Photo-ionization from young, hot stars is the main excitation mechanism for \Fe{} and hydrogen, whereas shocks are responsible for the \Htwo{} excitation. Our observations are broadly comparable with simulations where a Toomre-unstable, self-gravitating gas disk triggers a burst of star formation, peaking after about 30 Myr and possibly cycling with a period of about 200 Myr.
\end{abstract}
\section{Introduction}
\subsection{Active Galactic Nuclei and Star Formation}
Active galactic nuclei (AGN) and their host galaxies have a close association, with the mass of the super-massive black hole (SMBH) that ultimately powers the activity being correlated with several galaxy properties; stellar velocity dispersion \citep{Ferrarese2000,Gebhardt2000a}, galactic bulge mass \citep{Kormendy1993}, K-band luminosity \citep{Kormendy1995,Graham2013} and the light profile (S\'{e}rsic index) \citep{Graham2001}. As the ``sphere of influence'' of the SMBH is tiny compared to the scale of the galaxy, feedback mechanisms must link the mutual mass growths together. These can be both negative, where AGN energetics suppress star formation (SF) \citep{Puchwein2013} or positive, enhancing SF \citep[e.g.][]{VanBreugel1993, Mirabel1999,Mould2000}. Nearby galaxies hint that these two processes are not mutually exclusive, but are closely coupled \citep{Floyd2013,Rosario2010}. At high AGN powers, there is enough energy transfer to the interstellar medium (ISM) to suppress SF. However, for most of the time, AGN are in a low power mode; recent observations \citep{Villar-Martin2016} suggest that even luminous AGNs have low to modest outflows, not enough to suppress SF.Scenarios can be suggested where low-power outflows, radio jets or gravitational instabilities compress the ISM to enhance SF. \citet{Watabe2007} demonstrates a positive correlation between AGN and nuclear starburst luminosities; \citet{Zubovas2013} predicts this from numerical simulations. (See \citet{Watabe2007} for an overview.) Recent observations of the Phoenix Cluster with ALMA \citep{Russell2016} show that, even at very high powers, production of cold gas is stimulated by radio bubbles to provide SF fuel. 

Three-dimensional adaptive-mesh resolution (AMR) hydrodynamic simulation models can resolve the accretion disk, dusty torus, broad and narrow line regions and large scale in- and out-flows in a range of time and space scales \citep{Wada2009,Schartmann2009}. Observationally, with current instrumentation, we can resolve at sub-10 parsec scales those objects with a distance of less than 100 Mpc. 

\subsection{Our Program}
Our program is to study the nuclear activity of nearby early-type galaxies with radio emission for AGN and star-formation activity, to elucidate the details of the feedback mechanisms. We map the material flows though 2D velocity structures and study the gas excitation to determine the relative contributions of the activity modes. This mapping is at the smallest possible scale, close to the central engine. Our sample is based on the \citet{Brown2011} early-type  galaxy radio catalog, which demonstrated the likelihood that all massive early type galaxies harbor an AGN and/or have undergone recent star formation. Using this catalog, \citet{Mould2012} conducted long-slit spectroscopic observations and found about 20\% of objects showed IR emission lines; these have greater radio power for a given galaxy mass than those without such lines. We select elliptical and lenticular galaxies, as the fueling and stellar populations are likely to be simpler than spirals, stored gas is smaller and there is a minimum of nuclear obscuration. 

We also select objects with a distance $<$ 80 Mpc; resolving 2D gas flow velocity structures requires integral fields spectroscopes/units (IFS/IFU), and this distance limit enables adaptive optics (AO)-corrected IFU observations at sub-10 parsec resolution. AO performs best in the infrared, which also has the advantage of penetrating obscuring gas and dust to directly observe the nuclear region. Currently, there are 3 instruments that meet the combined requirements of a telescope in the 8-10m class, adaptive optics and a near infra-red IFU; these are SINFONI on  VLT \citep{Eisenhauer2003}, NIFS on Gemini North \citep{McGregor2003} and OSIRIS on Keck I \citep{Larkin2006}.

This paper is the second on the \citet{Mould2012} galaxies, the first was on NGC~2110 \citep{Durre2014}, which showed young, massive star clusters embedded in a disk of shocked gas from stellar winds which feed the black hole; the star formation rate is 0.3 \msun{} yr$^{-1}$, easily sufficient to produce the clusters on a million year timescale. The gas kinematics produced an estimate for the enclosed central mass of $3.2–-4.2 \times 10^{8}~\mmsun{}$.

Other work has concentrated on ``classical'' Seyferts, which normally reside in spirals. \citet{Riffel2015} summarizes the work of the AGNIFS group at the Universidade Federal de Santa Maria and the Universidade Federal do Rio Grande do Sul, Brazil, on 10 Seyfert AGNs; they find that for LINERs (Low-Ionization Nuclear Emission-line Region) and low-power Seyferts, gas in different phases has distinct distributions and kinematics; with ionized outflow rates in the range 10\pwr{-2} -- 10 \msun{} yr\pwr{-1} (in ionized cones or compact structures), and \Htwo{} inflow rates of 10\pwr{-1} -- 10 \msun{} yr\pwr{-1}. The stellar kinematics of Seyfert galaxies reveals cold nuclear structures composed of young stars, usually associated with a significant gas reservoir. \citet{Hicks2013} and \citet{Davies2014}, in their comparison of 5 active and 5 quiescent galaxies, reach the same conclusions; further finding that the quiescent galaxies have chaotic dust morphologies and counter-rotating molecular gas.

\citet{Davies2007a} observes moderately recent (10--300 Myr) starbursts around all 9 of their heterogeneous sample of AGNs, deducing episodic periods of star formation. They posit a 50--100 Myr delay between SF and AGN activity onsets, concluding that OB stars and supernova produce winds that have too high velocities to feed the AGN, but that evolved AGB stellar winds with slow velocities can be accreted efficiently onto the SMBH.

This paper investigates the link between AGN activity and star formation at small scales. We combine the near IR observations with supplementary optical IFU observations. This paper uses Vega system magnitudes and the standard cosmology of H$_0 = $73 \kms, $\Omega_{Matter}=0.27$ and $\Omega_{Vacuum}=0.73$.

\section{IC~630 as a Starburst}
IC~630 shows the second strongest IR emission line flux in the \citet{Mould2012} spectroscopic observation program (after the well-studied Seyfert 1 galaxy UGC3426/Mrk 3).

IC~630's basic details are from NASA/IPAC Extragalactic Database (NED)\footnote{\url{http://ned.ipac.caltech.edu/}} unless otherwise noted. IC~630 (Mrk 1259) has a type of S0 pec (morphological type -2) from the RC3 catalog \citep{DeVaucouleurs1991} with a redshift of 0.007277. Using the Virgo+GA+Shapley Hubble flow model, this gives a distance of 33.3 $\pm$ 2.3 Mpc, with a distance modulus of 32.6 mag (a flux-to-luminosity ratio of 1.33 $\times 10^{53}$ in cgs units, i.e. \ecs{} to erg s\pwr{-1} at the distance of IC~630).  It has starburst type activity \citep{Balzano1983}, rather than the classical AGN-type high-excitation emission lines of Seyfert galaxies; in fact, it is classified as a ``Wolf-Rayet'' galaxy with a super-wind, similar to M82 \citep{Ohyama1997}, with a high ratio of WR to O-type stars ($\sim$ 9\%) The outflow is seen almost face-on, with the estimated velocity of $\sim$ 710 km s\pwr{-1}. Strong optical emission lines of hydrogen,  [\ion{O}{3}] and [\ion{N}{2}] are seen, as well as \ion{N}{3}, \ion{N}{5}, \ion{He}{1} and \ion{He}{2}.

\begin{figure*}[!]
	\centering
	\includegraphics[width=1\linewidth]{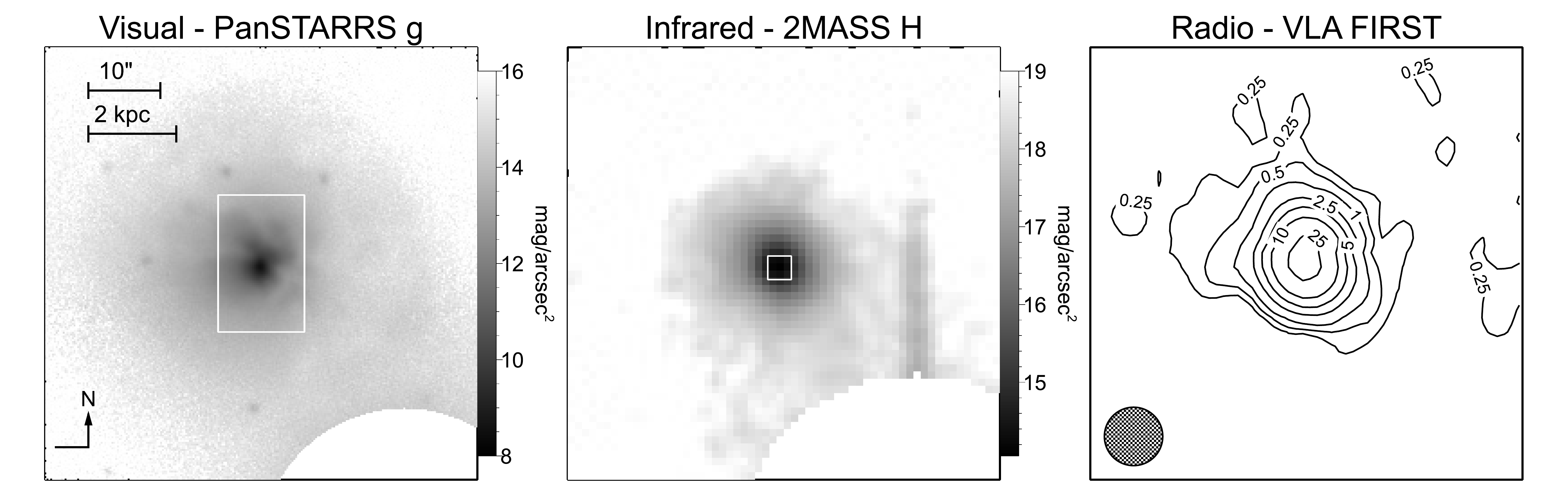}
	\caption{Left panel: optical image (PanSTARRS \textit{g}). Middle panel: infrared image (2MASS H). Right panel: radio (VLA FIRST 1.4 GHz) image.  The optical and IR  units are magnitude per square arcsec. The contoured radio flux units are mJy; the circle in the bottom left of the plot is the beam size. The optical image shows the field of view ($25\times38\arcsec$) for the optical WiFeS IFU as a rectangle (the regularly spaced dots are an image artifact). The IR image also shows the field of view for the infrared IFU observations ($3.3\times3.3\arcsec$) as a small white square. The image sizes are $60 \times 60\arcsec$, equivalent to $\sim$10 kpc.}  
	\label{fig:IC630_Images_VIS_IR_VLA}
\end{figure*}

To illustrate global morphology and the relevant scales, Fig. \ref{fig:IC630_Images_VIS_IR_VLA} shows images for this object; the optical from the PanSTARRS \textit{g} band image from the MAST PanSTARRS image cutout facility\footnote{\url{https://archive.stsci.edu/}}, the H band near-infrared from the 2MASS catalog \citep{Skrutskie2006} and the radio  at 1.4 GHz from the VLA FIRST survey \citet{Becker1995} image cutout facility\footnote{\url{http://third.ucllnl.org/cgi-bin/firstcutout}}. The IR image is somewhat confused by the diffracted light from the nearby star HD92200, which has been masked out.   The scale and orientation are the same in all images as shown on the optical image. The infrared image show a featureless spheroid with a published (NED) ellipticity of about 0.5 at a position angle (PA) of about 60$\degr$. In the PanSTARRS \textit{g} optical image (Fig. \ref{fig:IC630_Images_VIS_IR_VLA}), the nucleus is off-center with respect to the disk, and shows dust lanes crossing SW to NE across the nucleus, with a suggestion of a shell structure (most prominent in the NE quadrant). The radio image has a hint of a lobe towards the NW.

An estimate of the galaxy mass was calculated from the Spitzer Heritage Archive 3.6\um{} and 2MASS H band images. This yields $\sim 5.1 \times 10^{10}$ \msun{} (assuming a mass-to-light ratio of 0.8 at 3.6\um) and $\sim 6.4 \times 10^{10}$ \msun{}  from the 2MASS image (assuming a M/L of 1 at H).

\begin{figure}[!]
	\centering
	\includegraphics[width=1\linewidth]{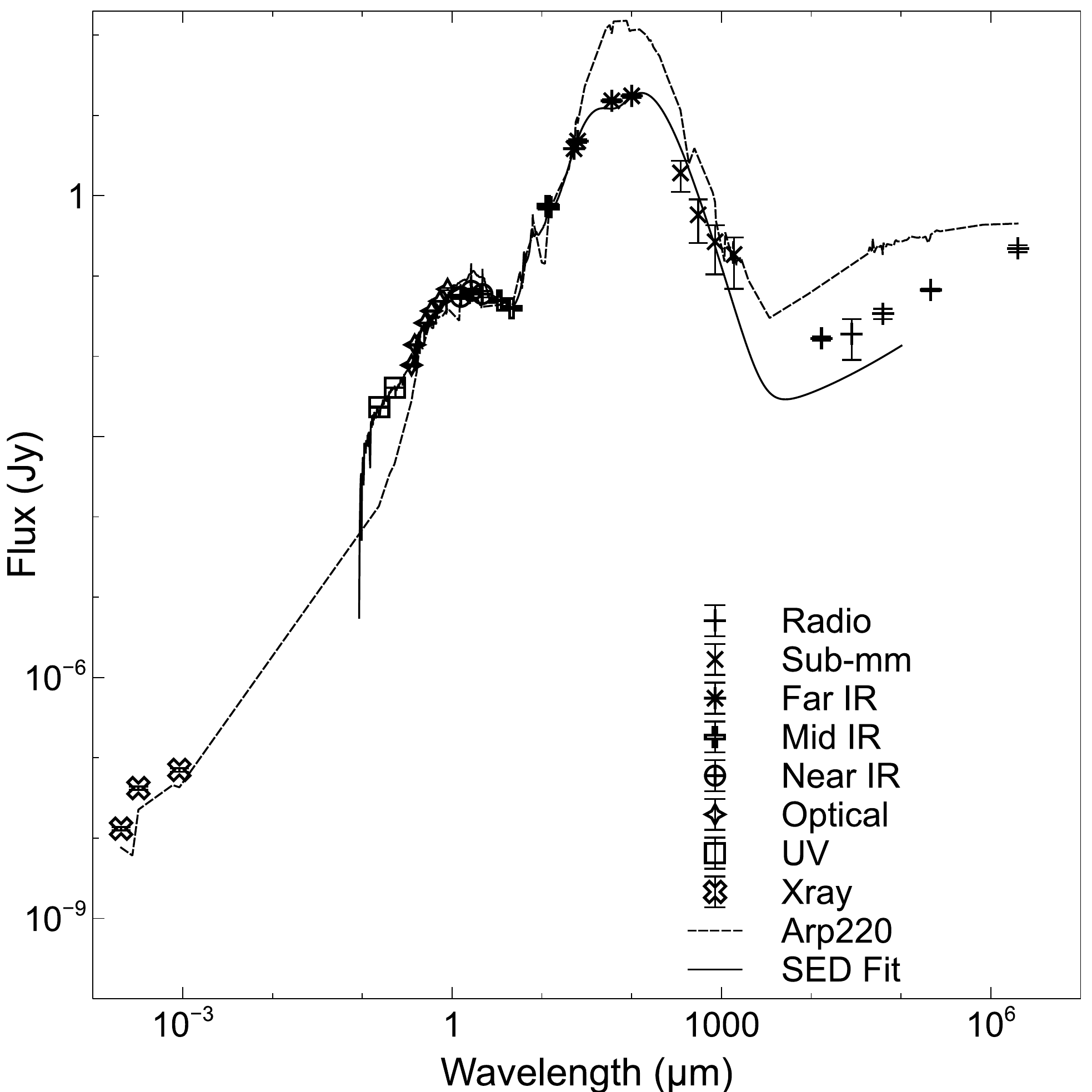}
	\caption{Photometric data from table \ref{tbl:Photometry} with \texttt{magphys} fit and example starburst SED (Arp220).}
	\label{fig:IC630_Photometry}
\end{figure}

To confirm the starburst characterization, the spectral energy distribution (SED) was compiled from existing data sources (table \ref{tbl:Photometry}), which is plotted in Fig. \ref{fig:IC630_Photometry}. This confirms the type, with the major peak at around 100\um~($\approx$ 30 K) being from dust heated by star formation. For comparison, the SED of the well-known starburst galaxy Arp 220 is also plotted, showing very similar features. The SED was also fitted using the \texttt{magphys} package with the HIGHZ extension to fit the Planck observations \citep{DaCunha2008,DaCunha2015}; the fit estimates the galaxy mass at $\sim1.5 \times 10^{10}~\mmsun$ (somewhat lower then the mass estimate from the near IR photometry) and a SFR of $\sim1.1~\mmsun~yr^{-1}$, in line with the values in table \ref{tbl:SFRs}, which presents derived SFRs from various flux indicators. It is noted that the radio flux is about a factor of 4 above the fit; this could be as a result of the uncertainties associated with the fit, which uses a prescription based on the far IR and radio correlation (da Cunha, private comm.) or from a highly obscured AGN. Nuclear starbursts are usually the result of galaxy interactions \citep{Bournaud2011} and the optical image shows a disturbed morphology.  An examination of images from various surveys around IC~630 does not reveal any candidate interacting galaxy, so we suggest that the starburst is generated by a minor merger.

\begin{table}[!]
	\centering
	\caption{IC630 Photometry}
	{\scriptsize 	\begin{tabular}{lccccc}
			\hline
			Band    &  Freq.   & $\lambda$ &   Flux   & e\_Flux  & Ref \\
			&   (Hz)   &   (\um)   &   (Jy)   &   (Jy)   &  \\ \hline
			Radio   & 1.50E+08 & 2.00E+06  & 2.20E-01 & 2.24E-02 &  1  \\
			& 1.40E+09 & 2.14E+05  & 6.70E-02 & 2.50E-03 &  2  \\
			& 4.77E+09 & 6.29E+04  & 3.40E-02 & 5.00E-03 &  3  \\
			& 1.06E+10 & 2.82E+04  & 1.90E-02 & 1.00E-02 &  3  \\
			& 2.31E+10 & 1.30E+04  & 1.67E-02 & 1.00E-03 &  3  \\
			Sub-mm  & 2.17E+11 & 1.38E+03  & 1.85E-01 & 1.16E-01 &  4  \\
			& 3.53E+11 & 8.50E+02  & 2.67E-01 & 1.62E-01 &  4  \\
			& 5.45E+11 & 5.50E+02  & 5.78E-01 & 3.20E-01 &  4  \\
			& 8.57E+11 & 3.50E+02  & 1.92E+00 & 8.07E-01 &  4  \\
			Far IR  & 3.00E+12 & 1.00E+02  & 1.75E+01 & 8.77E-01 &  5  \\
			& 5.00E+12 & 6.00E+01  & 1.52E+01 & 7.61E-01 &  5  \\
			& 1.20E+13 & 2.50E+01  & 4.78E+00 & 2.39E-01 &  5  \\
			& 1.32E+13 & 2.28E+01  & 3.85E+00 & 2.81E-02 &  6  \\
			Mid IR  & 2.50E+13 & 1.20E+01  & 7.20E-01 & 3.60E-02 &  5  \\
			& 2.59E+13 & 1.16E+01  & 7.39E-01 & 1.02E-02 &  6  \\
			& 6.52E+13 & 4.60E+00  & 3.95E-02 & 6.64E-04 &  6  \\
			& 8.96E+13 & 3.35E+00  & 4.93E-02 & 1.00E-03 &  6  \\
			Near IR & 1.38E+14 & 2.17E+00  & 5.96E-02 & 5.33E-03 &  7  \\
			& 1.83E+14 & 1.64E+00  & 6.53E-02 & 4.25E-03 &  7  \\
			& 2.40E+14 & 1.25E+00  & 5.51E-02 & 3.00E-03 &  7  \\
			Visual  & 3.37E+14 & 8.90E-01  & 6.86E-02 & 2.09E-03 &  8  \\
			& 4.11E+14 & 7.30E-01  & 4.85E-02 & 6.96E-04 &  8  \\
			& 5.08E+14 & 5.90E-01  & 3.67E-02 & 9.20E-04 &  8  \\
			& 6.00E+14 & 5.00E-01  & 2.60E-02 & 4.89E-04 &  8  \\
			& 7.83E+14 & 3.83E-01  & 1.39E-02 & 2.21E-04 &  8  \\
			& 8.50E+14 & 3.53E-01  & 7.76E-03 & 1.95E-04 &  8  \\
			UV      & 1.30E+15 & 2.31E-01  & 4.07E-03 & 2.99E-05 &  9  \\
			& 1.95E+15 & 1.54E-01  & 2.32E-03 & 4.01E-05 &  9  \\
			X ray   & 3.27E+17 & 9.18E-04  & 7.06E-08 & 3.53E-09 & 10  \\
			& 9.35E+17 & 3.21E-04  & 4.19E-08 & 2.10E-09 & 10  \\
			& 1.46E+18 & 2.06E-04  & 1.31E-08 & 6.55E-10 & 10  \\ \hline
	\end{tabular}}
	\tablerefs{{\scriptsize 1=MWA Gleam \citet{Hurley-Walker2016}, 2=VLA Sky Survey \citet{Condon1998}, 3=\citet{Bicay1995}, 4=Planck \citet{Adam2016}, 5=IRAS-S \citet{Kleinmann1986}, 6=WISE \citet{Wright2010}, 7=2MASS \citet{Skrutskie2006}, 8=Skymapper \citet{Wolf2016}, 9=GALEX - MAST at STSCI, 10=ASCA \citet{Ueda2005}}}
	\label{tbl:Photometry}
\end{table}
\begin{table*}[!]
\centering
\caption{Star formation rates for various indicators.}
\label{tbl:SFRs}
{\footnotesize 		\centering
\begin{tabular}{llrrrlrlc}
	\hline
	Passband  & Instrument  &      $\lambda/\nu/eV$ & Ref. &                  Flux & Units &          Luminosity & Units         &      SFR (\msun yr\pwr{-1})       \\ \hline
	Radio     & MWA-GLEAM   &               150 MHz &    3 &                  0.22 & Jy    & 2.1$\times 10^{22}$ & W Hz\pwr{-1}  &      2.8       \\
	Radio     & VLA         &               1.4 GHz &    2 &                 0.067 & Jy    & 8.9$\times 10^{21}$ & W Hz\pwr{-1}  &      3.6       \\
	Mid-IR    & WISE W4     &              22.8 \um &    1 &                   3.9 & Jy    & 6.8$\times 10^{43}$ & erg s\pwr{-1} &      10.3      \\
	Mid-IR    & WISE W3     &              11.6 \um &    1 &                   0.7 & Jy    & 2.4$\times 10^{43}$ & erg s\pwr{-1} &      5.8       \\
	H$\alpha$ & Oyakama NCS & 6562.8 $\textrm{\AA}$ &    5 & 2.77$\times 10^{-12}$ & \ecs  & 3.7$\times 10^{41}$ & erg s\pwr{-1} &      2.0       \\
	Far UV    & GALEX       & 1538.5 $\textrm{\AA}$ &    4 &  2.20$\times 10^{-3}$ & Jy    & 5.8$\times 10^{42}$ & erg s\pwr{-1} &      2.9       \\
	X-ray     & ASCA        &             0.7-2 keV &    6 & 2.30$\times 10^{-13}$ & \ecs  & 3.1$\times 10^{40}$ & erg s\pwr{-1} & 6.8$~\dagger$  \\
	X-ray     & ASCA        &            0.7-10 keV &    6 & 3.90$\times 10^{-13}$ & \ecs  & 5.2$\times 10^{40}$ & erg s\pwr{-1} & 13.0$~\dagger$ \\ \hline
\end{tabular}
}
\tablerefs{{\footnotesize 1=IRSA AllWISE Catalog \citep{Wright2010}, 2=NRAO VLA Sky Survey, \citet{Condon1998}, 3=MWA GLEAM Survey \citep{Hurley-Walker2016}, 4=GALEX data archive, 5=\citet{Ohyama1997}, 6=\citet{Ueda2005}.}}
\tablecomments{{\footnotesize Star formation rate indicators are from Brown et al. (2017, in preparation), except $\dagger$ from \citet{Ranalli2003}}}
\end{table*}
\section{Observations and Data Reduction}
\subsection{Observations}
\label{sec:Observations}
Our infrared observations were taken with NIFS on Gemini North in queue service observing mode and SINFONI on VLT-U4 (Yepun) in classical/visitor observing mode. Observations were carried out using adaptive optics with laser guide stars, as per table \ref{tbl:ObservationLog}.

Each dataset consists of two 300 second observations, combined with a sky frame of 300 seconds, in the observing mode ``Object-Sky-Object". The NIFS observations used simple nodding to the sky position, which was 30\arcsec~in both RA and Dec; for SINFONI the offset was 30\arcsec{} in Dec, plus a 0.05\arcsec{} jittering procedure.
\begin{table*}[!]
\centering
	\caption{Observation Log}
	\footnotesize
	\begin{tabular}{llrlllr}
		\hline
		Filter  & Date                                     &                  Exp. Time & Spaxel Resolution           & Program ID    & Standard Star     &       PSF FWHM (pixel)\tablenotemark{e} \\
		        & Wavelength                               &                   Datasets & Seeing\tablenotemark{c}     &               & mag               & PSF Res. (\arcsec) \\
		        & $\Delta\lambda/\lambda\tablenotemark{a}$ & $\Delta$V\tablenotemark{b} & Airmass                     & Instrument    & Eff. Temperature  &      PSF Res. (pc) \\ \hline\hline
		J       & 11-Apr-2014                              &                      600 s & 100 mas                     & ESO           & HIP55051 (B2/B3V) &                    4.5 \\
		        & 1088--1412 nm                            &                          1 & 1.1\arcsec\tablenotemark{d} & 093.B-0461(A) & 7.88              &                  0.225 \\
		        & 2360                                     &                        127 & 1.07                        & SINFONI       & 24000             &                     37 \\ \hline
		H       & 6-May-2015                               &                     1800 s & 103 x 40 mas                & Gemini        & HD88766 (A0V)     &                    4.5 \\
		        & 1490--1800 nm                            &                          3 & 0.5\arcsec                  & GN-2015A-Q-44 & 7.63              &                  0.225 \\
		        & 5300                                     &                         57 & 1.14                        & NIFS          & 12380             &                     37 \\ \hline
		K       & 5-May-2015                               &                     1800 s & 103 x 40 mas                & Gemini        & HD88766 (A0V)     &                    5.2 \\
		        & 1990--2400 nm                            &                          3 & 0.7\arcsec                  & GN-2015A-Q-44 & 7.62              &                   0.26 \\
		        & 5300                                     &                         57 & 1.16                        & NIFS          & 12380             &                     43 \\ \hline
		Optical & 18-Nov-2016                              &                      600 s & 1 x 2 sec                   & ANU 2.3 m     & HD88766 (A0V)     &  \\
		        & 350--900 nm                              &                          1 & 2\arcsec                    & 4160069       & 7.90 (V)          &  \\
		        & 3000                                     &                        100 & 1.55                        & WiFeS         &                   &  \\ \hline
	\end{tabular}
	\tablenotetext{a}{{\footnotesize ~Instrumental spectral resolution}}
\tablenotetext{b}{{\footnotesize ~Velocity resolution (\kms)}}
\tablenotetext{c}{{\footnotesize ~Seeing from observer's report (SINFONI and WiFeS) or Mauna Kea Weather Center DIMM archive (NIFS)}}
\tablenotetext{d}{{\footnotesize ~From the collapsed data cubes of the standard stars, the seeing was probably somewhat better; $\sim$0.5\arcsec}}
\tablenotetext{e}{{\footnotesize ~PSF estimate described in section \ref{sec:TCFCPE}}}
\label{tbl:ObservationLog}
\vspace{0.5cm}
\end{table*}
Ancillary calibration observations were carried out on each night. For NIFS, these consist of arc, flat field, dark and Ronchi slit-mask  (for spatial calibrations). For SINFONI, these are dark current, flat field, linearity and distortion flat fields, arc and arc distortion frames. These are combined with static calibration data; line reference table, filter dependent setup data, bad pixel map and atmospheric refraction reference data.

Our optical observations were taken on  using the WiFeS instrument \citep{Dopita2010,Dopita2007} on the Australian National University’s 2.3 m telescope at Siding Spring Observatory. The WiFeS IFU has a $25\arcsec\times38\arcsec$ field of view and $1\arcsec\times1\arcsec$ spaxels. The B3000 (3500–-5800\AA) and R3000 (5300–-9000\AA) gratings were used along with the RT560 dichroic. The instrument was used in ``Classical Equal'' observation mode, with an average seeing of 2\arcsec.

\subsection{Data Reduction}
For the NIFS observation sets, the standard Gemini \texttt{IRAF} recipes\footnote{\url{http://www.gemini.edu/sciops/instruments/nifs/data-format-and-reduction}} were followed. This consists of creating baseline calibration files, then reducing the object and telluric observations using these calibrations. Each on-target frame is subtracted by the associated sky frame, flat fielded, bad pixel corrected and transformed from a 2D image to a 3D cube using the spatial and spectral calibrations. The resulting data cubes are spatially re-sampled to 50$\times$50 mas pixels. The resulting 6 data cubes for each filter set were manually registered by collapsing the cube along the spectral axis, measuring the centroid of the nucleus, re-centering each data cube and average combining them.

For the SINFONI observation sets, we used the recommendations from the ESO SINFONI data reduction cookbook and the \texttt{gasgano}\footnote{\url{http://www.eso.org/sci/software/gasgano.html}} software pipeline (version 2.4.8). Bad read lines were cleaned from the raw frames using the routine provided in the cookbook. Calibration frames were reduced to produce non-linearity bad pixel maps, dark and flat fields, distortion maps and wavelength calibrations. Sky frames are subtracted from object frames, corrected for flat field and dead/hot pixels, interpolated to linear wavelength and spatial scales and re-sampled to a wavelength calibrated cube, which also have pixels of $50\times50$ mas size. The reduced object cubes are mosaicked and combined to produce a single data cube.

The spectra were not reduced to the rest frame, as the target lines have good signal-to-noise (S/N), making identification unproblematic. The three final infrared data cubes (one each for J, H and K band) conveniently all have the same native spatial sampling (0.05\arcsec); these were resized so all were 66$\times$66 pixels (3.3$\times$3.3\arcsec), and re-centered to the brightest pixel (the nuclear core), with a field of view of $\sim$540$\times$540 pc at the galaxy. With this resolution, the plate scale is 8.2 pc pixel\pwr{-1}. 

The optical WiFeS data were reduced in the standard manner using the PyWiFeS reduction pipeline of \citet{Childress2013}, with flat-fielding, aperture and wavelength calibration, and flux calibration from the standard star. The spatial pixel scale of 1\arcsec{} is equivalent to 164 pc; the whole field of view is 4.1 $\times$ 6.6 kpc. The data reduction produces a data cube for each of the blue and red filters. These were attached together in the  wavelength axis and the red cube re-sampled to the same dispersion as the blue cube (0.0774 nm pixel\pwr{-1}). The spectrum at each pixel showed considerable sky background, including skyline emission, especially red-ward of 7200 \AA; this background was removed by subtracting the median spectrum of purely sky pixels. 

\subsection{Telluric Correction, Flux Calibration and PSF Estimation}
\label{sec:TCFCPE}
The telluric and flux calibration data reduction was carried out for each infrared instrument in the same manner. The standard stars (listed in table \ref{tbl:ObservationLog}) were used for both telluric correction and flux calibration. The telluric spectrum can be modeled by a black-body curve for the appropriate temperature plus simple Gaussian fits to remove the hydrogen and helium lines. The best blackbody temperature fit to the telluric star's spectrum observed in the infrared is somewhat higher than the stellar type's nominal optical temperature (about a factor of 1.3 times); the hydrogen opacity is lower at infrared wavelengths, so we observe a lower (and therefore hotter) layer in the stellar atmosphere.  This was confirmed by checking the stellar atmospheric model templates from \citet{Castelli2004} (available from the Space Telescope Science Institute\footnote{\url{ftp://ftp.stsci.edu/cdbs/grid/ck04models/}}), using the \texttt{ckp00} set of models for standard main sequence stars of solar metallicity.

To extract the spectrum of the star, the aperture to be used must be carefully determined. With a multiple elements in the optical train, the IFU image shows significant scattered light over more than 1 second radius. The aperture is set manually from the cube median image, using logarithmic scaling. A region outside this aperture was chosen to set the background level offset. For telluric correction,  a black-body curve of the appropriate temperature is removed and the absorption lines interpolated over. This was then normalized and divided into each spectral element of the science data cube. 

Flux calibration was done by reference to the spectrum and the J, H or K magnitude from the 2MASS catalog. The magnitude is converted to flux using the Gemini Observatory calculator ``Conversion from magnitudes to flux, or vice-versa''\footnote{\url{http://www.gemini.edu/sciops/instruments/midir-resources/imaging-calibrations/fluxmagnitude-conversion}} \citep{Cohen1992a}. The count calibration is done by averaging over a 10 nm range around the filter effective wavelengths (J 1235nm, H 1662nm, K 2159nm)  to get the counts and the flux calibration is computed (in \ecs{} count\pwr{-1}). These effective wavelengths will be used subsequently for surface brightness maps, along with the Johnson V effective wavelength (550nm).

\begin{figure}[!]
	\centering
	\includegraphics[width=1\linewidth]{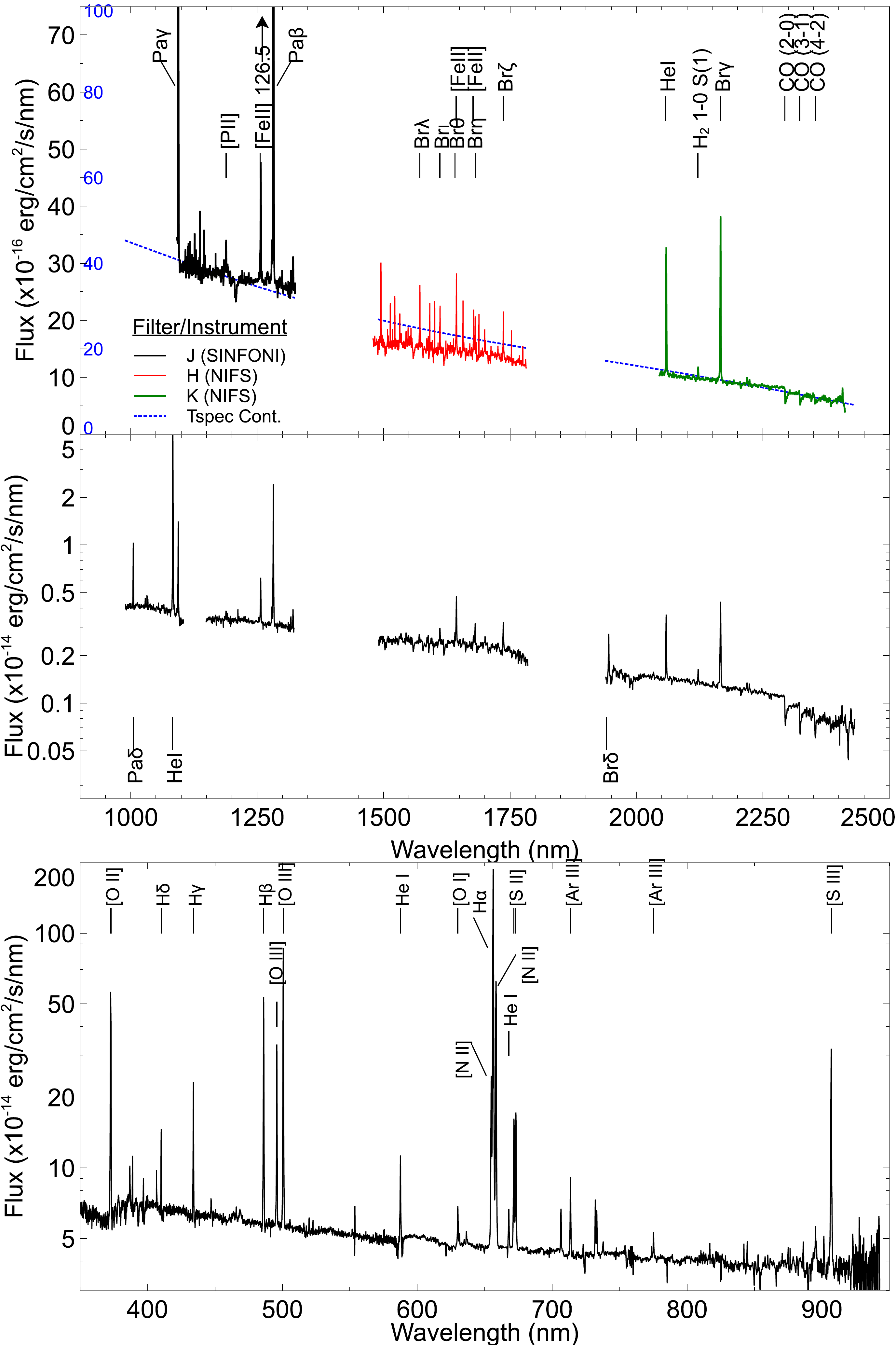}
	\caption{IC~630 Spectra. Top panel: IFU spectra of nuclear region, with Triplespec smoothed continuum. The main nebular emission lines are marked, as well as the molecular CO absorption band-heads in the K band. The vertical axis scales are: black numbers - IFU observations, blue numbers - Triplespec smoothed continuum. Middle panel: Triplespec spectrum; the flux is plotted on a log scale to encompass the large dynamic range. Extra nebular lines that are out of the IFU spectral ranges are also marked. Bottom panel: WiFeS optical spectrum, with main emission lines marked.}
	\label{fig:IC630_Spectrum}
\end{figure}

It is well known that the flux calibration for IFU instruments can produce uncertainties of the order of 10\%; our observations are also from 2 different instruments at three separate dates. We therefore cross-checked the calibration against the long-slit infrared observations from \citet{Mould2012}. Using the \textit{longslit} function in the data viewer and analysis package \texttt{QFitsView} \citep{Ott2016}, we extracted the spectra of a pseudo-longslit from each data cube with the same width and position angle (1\arcsec=20 pixels, PA=208\arcdeg) and compared it with the \citet{Mould2012} long-slit observations. These were plotted together as shown in Fig. \ref{fig:IC630_Spectrum}; the three IFU observations are smoothly continuous, demonstrating good flux calibration. These are plotted against a polynomial fit of the Triplespec continuum. The Triplespec flux is somewhat higher then the IFU fluxes ($\sim 25 \%$); the long slit takes in more disk light than the 3.3\arcsec{} IFU FOV, plus there is more scattered light in the IFU optical train. The spectral slopes are in good agreement; however the H band flux is somewhat lower than expected. This is probably due to flux calibration uncertainties. The optical spectrum of the central 3.3\arcsec{} from the WiFeS data cube is also plotted; the optical and IR continua are also smoothly continuous.

The point spread function (PSF) for the instruments was estimated from the standard star observations by fitting a 2D gaussian to the collapsed standard star cube. For the SINFONI observations, the AO correction was not applied for the star, to prevent saturation; the PSF was estimated using an alternative star with a different spatial sampling, observed on the same night. The resulting gaussian fits are somewhat elliptical, we use the major axis FWHM. The results listed in table \ref{tbl:ObservationLog}, showing the FWHM PSF in pixels, angular and spatial resolution.
\subsection{Instrumental Fingerprint}
\begin{figure}[!]
	\centering
	\includegraphics[width=1\linewidth]{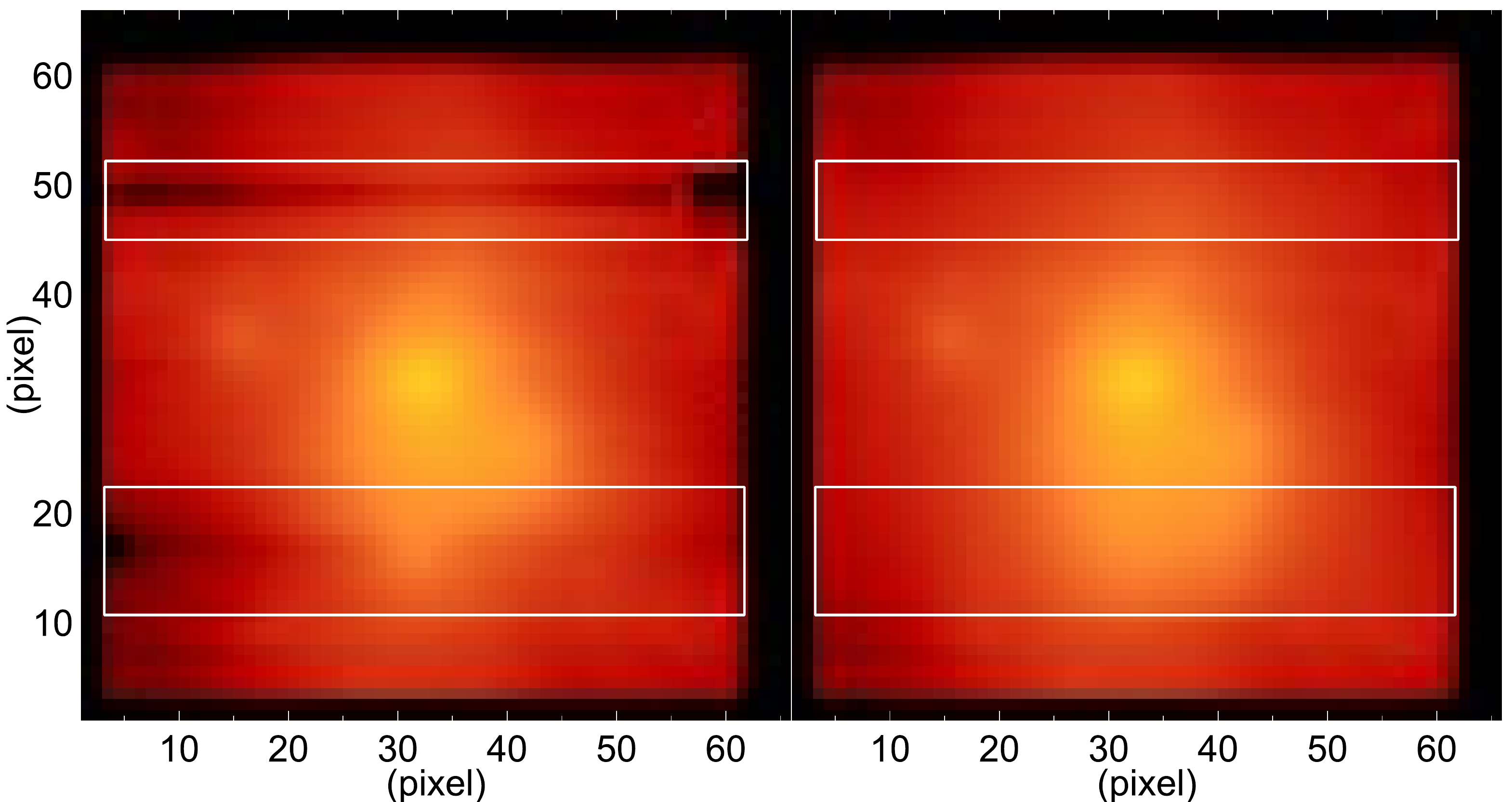}
	\caption{Left panel: SINFONI median image (log scale), showing instrumental ``fingerprints'' outlined. Right panel: After interpolation over the fingerprint.}
	\label{fig:IC630_J_Median}
\end{figure}

It is well-known that IFUs can have ``instrumental fingerprints'' that are not corrected by the standard calibration techniques. \citet{Menezes2014,Menezes2015a} demonstrate this for SINFONI and NIFS data cubes; they use the Principal Component Analysis (PCA) Tomography technique to characterize the fingerprints (which on SINFONI data cubes show broad horizontal stripes) and remove them. This is visible our data cubes; we median collapse all wavelengths and display on a logarithmic scale. In fact we see two horizontal stripes at y axis pixels 11-24 and 48--51, as shown in Fig. \ref{fig:IC630_J_Median}, and note this is different to the pattern found in \citet{Menezes2015a}. This fingerprint affects further results, especially for extinction measures. An attempt to apply the PCA technique to remove the fingerprints failed, as the fingerprint amplitude is comparable to the data and appeared strongly in the tomogram corresponding to eigenvector E1. As an alternative, we simply interpolated over the fingerprint in the y axis direction at each spectral pixel. The resulting data cube median is also shown in Fig. \ref{fig:IC630_J_Median}. 

\section{Results}
\subsection{Continuum Emission}
\begin{figure*}[!]
	\centering
	\includegraphics[width=1\linewidth]{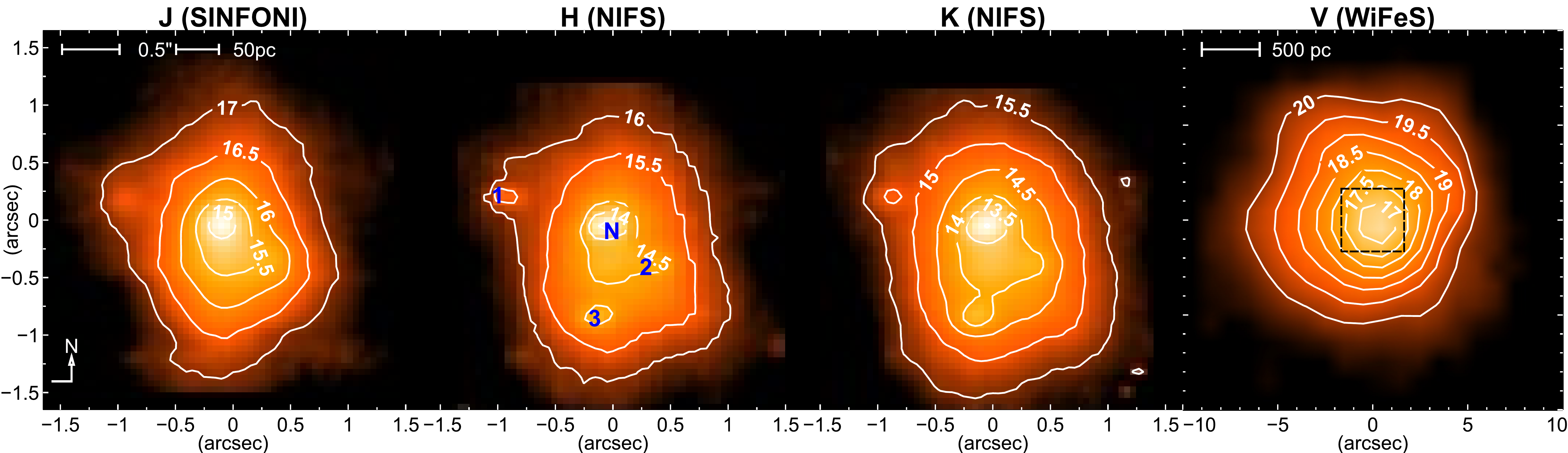}
	\caption{First, second and third panel: J, H and K band surface brightness. Labels (in blue) are the nuclear cluster/disk (`N'), secondary clusters (`1' and `3') and the ridge-like feature (`2'). The contour values are chosen to delineate the features. Fourth panel: Optical (V) surface brightness - note the change of spatial scale. The central square indicates the IR IFU fields of view. All values are in mag arcsec\pwr{-2}.}
	\label{fig:IC630_Continuum_JHK}
\end{figure*}

All image plots in this paper will use the same scale unless otherwise noted, i.e. $3.3 \times 3.3\arcsec$ a side FOV (540$\times$540 pc, 1 pixel = 8.2 pc), with North being up, and East to the left. For the WiFeS optical images, the plots are 20\arcsec~a side FOV ($3.3\times3.3$ kpc, 1 pixel = 164 pc). The RA and Dec values are given relative to the nuclear cluster center. Note the scale lengths of 0.5\arcsec{} and 50 pc. Fig. \ref{fig:IC630_Continuum_JHK} presents the stellar light around the nucleus, showing the individual J, H, K and V band surface brightness maps in units of magnitude per square arcsec. These were extracted from the respective data cubes by averaging over the 10 nm around filter effective wavelengths, dividing by the pixel area and converting to magnitude using the method described above.

In the IR, the nuclear region (labeled ``N") presents as a central clump with half-light radius of about 50 pc. There are 3 secondary features, a ridge at PA 240$\degr$ extending about 90 pc (labeled ``2'') and two local light maxima at PA/radius 73$\degr$/130pc and 175$\degr$/125pc (labeled ``1'' and ``3''). The ridge extension has a hint of a spiral structure. These features are more prominent in the H and K bands than the J band, which is consistent with greater dust penetration at longer wavelengths. The H band also has somewhat better observational resolution than the K band. The flux density for the nuclear cluster can be estimated by fitting a 2D Gaussian to the K band image (the least obscured data); this is 1.72 \ecsf{} nm\pwr{-1}. The Gemini Observatory flux to magnitude calculator gives a magnitude of 16.0 for the 2MASS K filter. At the distance magnitude of 32.6 and the solar K band magnitude of 3.28 \citep{Binney1998}, this gives a luminosity L = 9.0 $\times$ 10\pwr{7} \lsun{}. 

Comparing a Gaussian fit to each of the secondary features, the FWHM of each is comparable to or just marginally larger than that of the standard star; therefore we cannot say that these clusters are resolved.

\begin{figure}[!]
	\centering
	\includegraphics[width=1\linewidth]{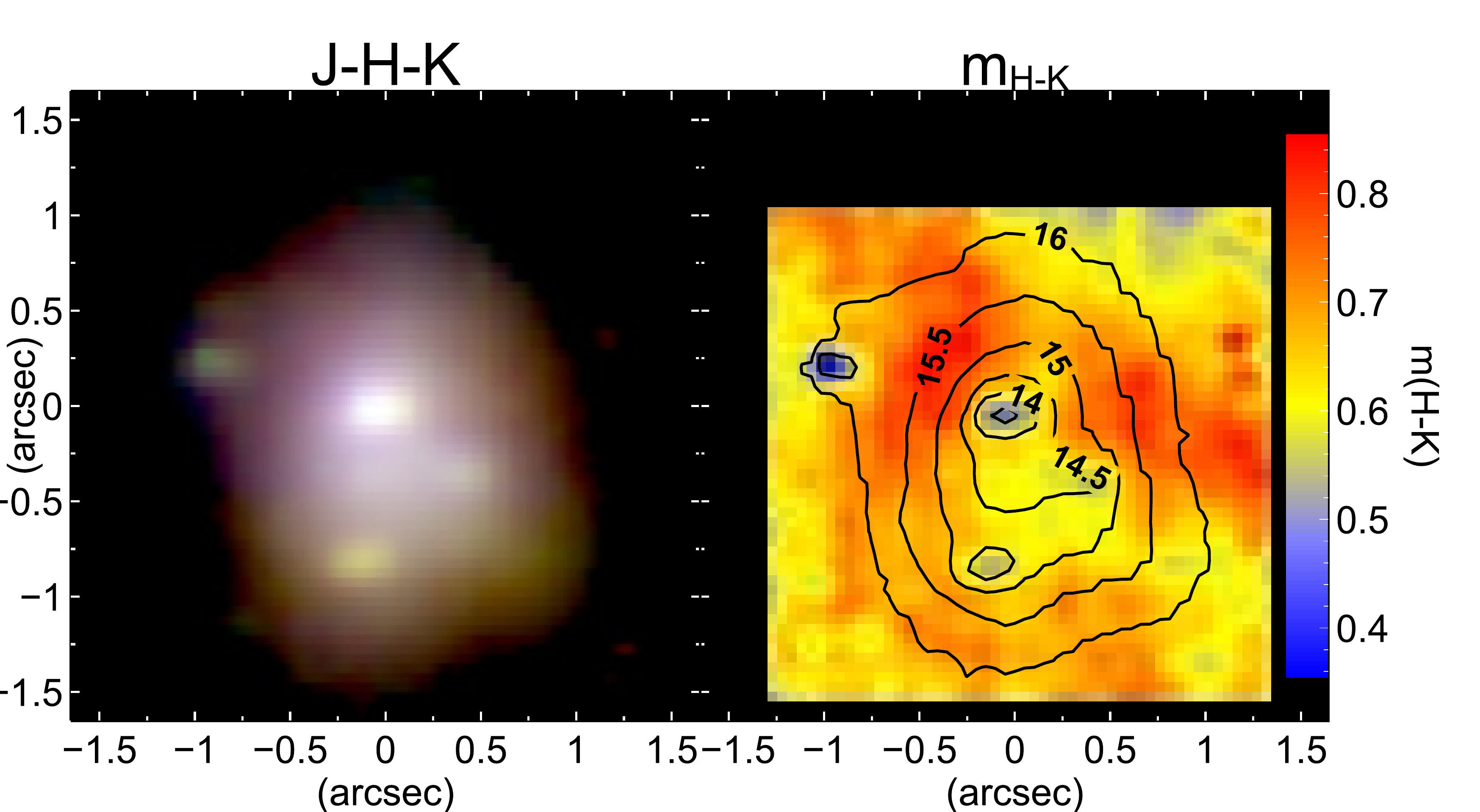}
	\caption{Left panel: Continuum 3 Color Image (J=blue, H=green, K=red). Right panel: H-K colour (in mag). The black contour plot is the H-band surface brightness, as per Fig. \ref{fig:IC630_Continuum_JHK} (second panel).}
	\label{fig:IC630_Continuum_3C}
\end{figure}

Stellar colors, indicative of population age and obscuration, are shown in the false-color continuum image (Fig. \ref{fig:IC630_Continuum_3C}, left panel), was created by layering the J, H and K filter total flux values (using blue, green and red colors, respectively). The individual images have been smoothed by a 2D Gaussian at 2 pixels width to remove pixelation; the colors have also been enhanced to bring out the salient features.  The right-hand panel of Fig. \ref{fig:IC630_Continuum_3C} shows the H--K magnitude; the J-band magnitude was not used, as the interpolation to remove the instrumental fingerprint obscures cluster ``3''. The H-band magnitude contours are over-plotted, with the same values as in Fig. \ref{fig:IC630_Continuum_JHK}; the nucleus and other cluster features have lower H--K magnitudes, indicative of bluer (younger) stellar populations. The optical surface brightness from the WiFeS data shows a featureless bulge; the  nucleus is offset like the PanSTARRS \textit{g} image.
\subsection{Stellar Kinematics}
\label{sec:StellarKinematics}
\begin{figure}[!]
	\centering
	\includegraphics[width=1\linewidth]{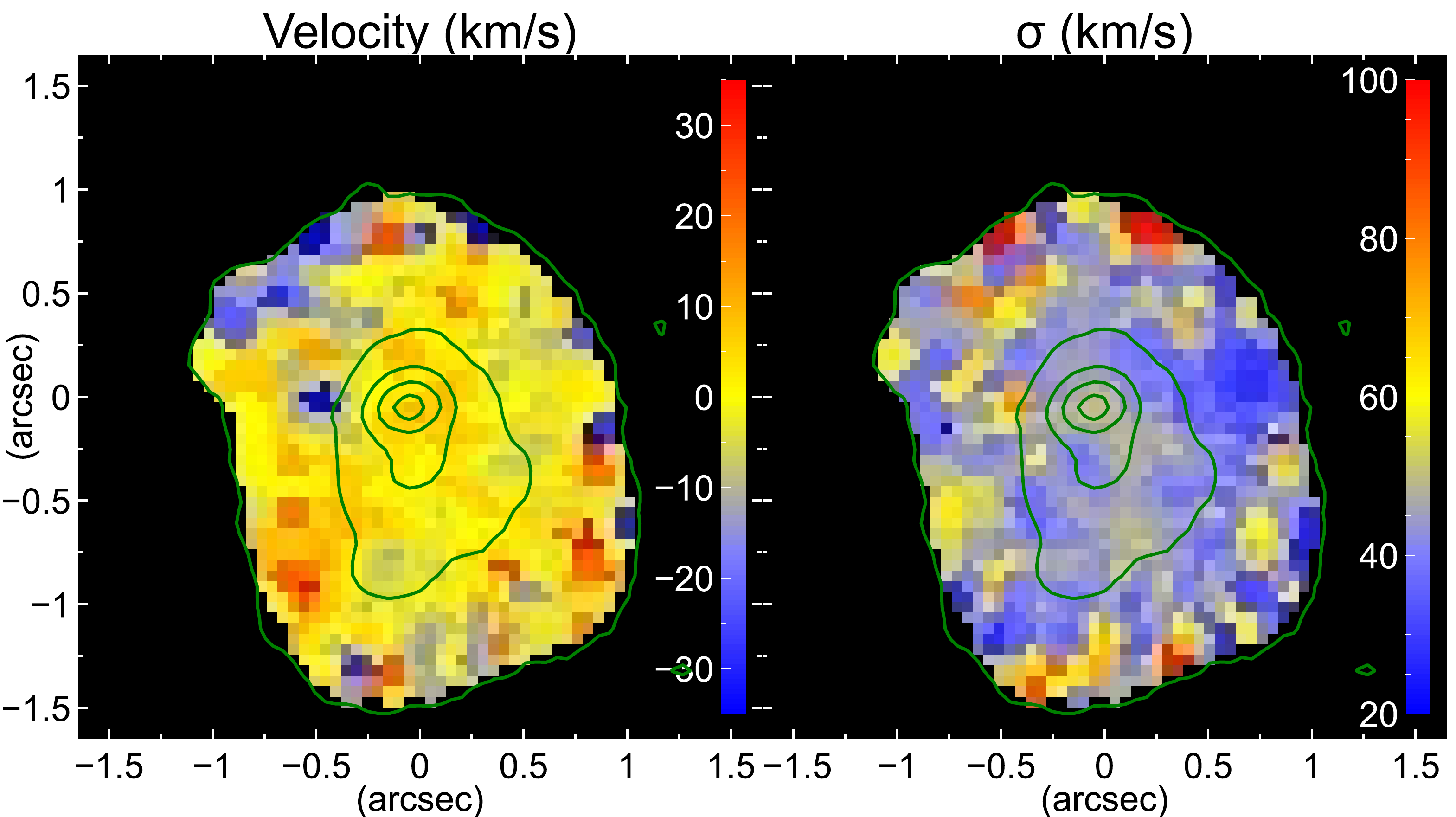}
	\caption{Stellar kinematics from CO band-heads. Left panel: velocity. Right panel: velocity dispersion. Values in \kms. The contours are the K-band flux, in the range 10 -- 90\% in steps of 20\% of the maximum value.}
	\label{fig:IC630_Stellar_Kinematics}
\end{figure}

The stellar kinematics were investigated using the CO band-heads in the range 2293--2355 nm, using the penalized pixel-fitting (pPXF) method of \citet{Cappellari2004}. The Gemini spectral library of near-infrared late-type stellar templates from \citet{Winge2009} was used, specifically the NIFS sample version 2. The observed K-band data cube was reduced to the rest frame (using z=0.007277, equivalent to a velocity of 2182 \kms) and normalized. Both the observations and template trimmed to the wavelength range of 2270--2370 nm. We followed the example code for kinematic analysis from the code website\footnote{\url{http://www-astro.physics.ox.ac.uk/\~mxc/software/\#ppxf}}. The ``Weighted Voronoi Tessellation'' (WVT) \citep{Cappellari2003} was used to increase the S/N for pixels with low flux; we use the \textit{voronoi} procedure in \texttt{QfitsView}. This aggregates spatial pixels in a region to achieve a common S/N. This needs both signal and noise maps; these are obtained from the fit and error of the stellar velocity dispersion value at each pixel calculated by the pPXF routine; in this case the S/N target was 250. Fig. \ref{fig:IC630_Stellar_Kinematics} displays the results. The velocity field has a range of $\pm$40 km s\pwr{-1} and shows no sign of ordered rotation. The zero value of the velocity was set as the median value returned from the pPXF code (31.5 \kms, compared to the instrumental velocity resolution of 57 \kms). The velocity dispersion range is 30--80 km s\pwr{-1}. 
Using the \mbh{}--$\sigma_{*}$ relationship from \citet{Graham2011} and using a velocity dispersion of 43.9 ($\pm$ 4.8) \kms{} (the average value and standard deviation of the central 1$\arcsec$ of the velocity dispersion map) and their relationship for elliptical galaxies (their table 2), we obtain $M_\bullet = 2.25~(-1.6, + 5.1) \times 10^{5}~\mmsun{}$. This should be regarded as an upper limit, as the relationship is derived only down to 60 \kms; however see \citet{Graham2016a} on the galaxy LEDA 87300. The relationship for early type galaxies in \citet{McConnell2013} gives $M_\bullet = 4.2~(-3.4, +1.8) \times 10^{4}~\mmsun{}$. The formulation in \citet{Kormendy2013} gives $M_\bullet = 3.9~(-2.4, +5.9) \times 10^{5}~\mmsun{}$. These rather wide error estimates indicate the possibility that there is no SMBH in the nucleus of this galaxy (within 1.4, 1.2 and 1.6 $\sigma$, respectively). If a BH of that mass exists, the sphere of influence is less that 1 pc.

To confirm this small SMBH size, we fitted a S\'{e}rsic index to the 2MASS K$_S$ image using the \texttt{Galfit3} application \citep{Peng2009}. This found an index of 0.824; using the relationship of \citet{Savorgnan2013}, we derive $M_\bullet = 1.14 (-0.25, +0.78) \times 10^{5} \mmsun{}$, compatible with the value derived from the stellar velocity dispersion.

Using the BH mass computed from the \citet{Graham2011} relationship, the X-ray luminosity, the relationship between X-ray and bolometric luminosity from \citet{Ho2009} ($ L_{Bol} = 16 \times L_{X}~(2-10~keV)$) and the Eddington luminosity ($L_{Edd} = 1.3 \times 10^{38}~M_{\bullet}/M_{\odot}~erg~s^{-1}$), we derive the Eddington ratio ($R_{Edd} = L_{Bol} /L_{Edd} = 1.4~(+3.3,-0.98) \times 10^{-2}$) and the corresponding accretion rate $\dot{M}_{acc} = \dfrac{L_{Bol}}{c^2 \eta} = 7.4 \times 10^{-5}~\mmsun~yr^{-1}$ using the standard efficiency factor $\eta$ of 0.1.
\subsection{Extinction}
The \textit{H-K} reddening map can be extended and quantified by using measures of extinction, where the known ratio of a pair of emission lines is compared against observations and extrapolated to the $\bv$ extinction and $A_V$ (the absolute extinction in the $V$ band). In general, the formula is as follows:
\begin{equation}
E_{B-V} = \alpha_{\lambda_1,\lambda_2} log\left( \dfrac{R_{\lambda_1,\lambda_2}}{F_{\lambda_1}/F_{\lambda_2}}\right) 
\label{eqn:Xtn}
\end{equation}
where $R_{\lambda_1,\lambda_2}$ is the intrinsic emissivity ratio of the two lines and $\alpha_{\lambda_1,\lambda_2}$ extrapolates from the emission line wavelengths to $\bv$. From the \citet{Cardelli1989} reddening law, $A_V=R_{V}\times E_{B-V}$ where $R_V$ is the extinction ratio. We use a value of $R_V$ = 3.1, the standard value for the diffuse ISM. Following the Cardelli parametrization of the reddening law in the infrared, we derive (where the wavelengths are in nm): 
\begin{equation}
\alpha_{\lambda_1,\lambda_2}=\frac{2.95\times10^{-5}}{(\lambda_2^{-1.61}-\lambda_1^{-1.61})}
\end{equation}
From our observations, we have several methods of deriving the extinction; the various ratios between hydrogen recombination lines, and between the \Fe{} emission lines at $\lambda$ = 1257 nm and 1644 nm. The values of $\alpha_{\lambda_1,\lambda_2}$ and $R_{\lambda_1,\lambda_2}$ are given in Table \ref{tbl:ExtinctionCompare} below for the individual line ratios.

The \HI{} ratios for (\Ha, \Hb, \pab, \pag{} and \brg) are determined by case B recombination and assuming an electron temperature $T_e = 10^4$ K and a density $n_e=10^3~cm^{-3}$ \citep{Hummer1987}. The intrinsic flux ratio is a weak function of both temperature and density. Over a range $5,000<T_{e}<10,000$ K and $100<n_e<1000~\text{cm}^{-3}$, this varies by only 5\%.

As the \Fe{} lines 1644 nm and 1257 nm share the same upper level, $a^{4}D\rightarrow a^{6}D$ and $a^{4}D\rightarrow a^{4}F$, respectively, their ratio is a function purely of the transition probabilities and wavelengths. The values of intrinsic emissivity ratio from the literature are discrepant; a value 1.36 is derived from \citet{Nussbaumer1988} Einstein coefficients, while the values from  \citet{Quinet1996}, give a value to 1.03, which decreases the derived $E_{B-V}$ value by 1 mag and the $A_V$ value by about 3 mag (see further discussion in \citet{Koo2016a}, \citet{Hartigan2004} and \citet{Smith2006}). Using the value of 1.36 brings the extinction derived from the \Fe{} ratio into line with that from the \HI{} ratio, so it will be adopted. The flux values are derived as described in section \ref{sec:GasFlux} below. The visual extinction is derived as above and the galactic foreground extinction of 0.16 mag \citep{Burstein1982} is subtracted from the derived extinction. The maps are presented in Fig. \ref{fig:IC630_Extinction_Maps}, showing the $A_{V}$ as derived from the various line ratios. The average, maximum and minimum values derived from the various line ratios are shown in table \ref{tbl:ExtinctionCompare}. For the IR spectral lines, the average values show close agreement, with some variances in the extrema; the plots show patchy extinction, but none of the methods line up spatially with each other. We hypothesise that this is caused by several effects: 
\begin{itemize}
	\item The \Fe{}  and \HI{} emissions come from gas in different excitation states which are not co-located and therefore have different LOS optical depths.
	\item The \pab/\brg{} and the \Fe{} emissions are measured from different data cubes with pixel to pixel variations both from alignment and calibration. These are also subject to the instrumental fingerprint mentioned above.
	\item The \pag/\pab{} ratios (which presumably do not suffer from calibration problems) are sensitive to flux measurement errors, which cause large variations since the wavelengths are close together. That noted, this ratio has the smallest variation over the whole field, being 2.2 mag. in $A_V$ (0.7 in $E_{B-V}$)
\end{itemize}
\begin{table}[]
	\centering
	\caption{Extinction calculation parameters for equation \ref{eqn:Xtn}. $A_V$ for the IR spectral lines is for a circle of 100 pc radius around the nucleus, with maximum and minimum values at 5 and 95 percentile to remove extrema. For the optical lines (\Ha/\Hb) the values are for the inner 3$\times$3 \arcsec.}
	\label{tbl:ExtinctionCompare}
	\begin{tabular}{lrrrrr}
		\hline
		                         &                                &                           &     & $A_V$ &  \\
		Spectral Lines           & $\alpha_{\lambda_1,\lambda_2}$ & $R_{\lambda_1,\lambda_2}$ & Avg &   Max & Min \\ \hline
		\pab/\brg                &                           5.22 &                      5.88 & 3.7 &   5.9 & 2.0 \\
		\Fel{}1257/$\lambda$1644 &                           8.21 &                      1.36 & 3.6 &   5.8 & 1.4 \\
		\pag/\pab                &                          10.32 &                      0.55 & 2.8 &   4.2 & 1.5 \\
		\Hb/\Ha                  &                           1.63 &                      0.35 & 0.8 &   1.6 & 0.2 \\
	\end{tabular}
\end{table}

Since there is no definitive pattern of extinction, we will use a single average value ($A_V = 3.4$) over the whole field to derive the extinction correction. Using the Cardelli reddening law, the ratio $A_{\lambda}/A_V$ is 0.287, 0.178 and 0.117 (an increase in surface brightness of 0.97, 0.60 and 0.40 magnitudes) for the J, H and K band filters respectively. The de-reddened H-K map values are reduced by only 0.2 magnitude.

Extinction measures derived from emission lines are usually higher than those for stellar populations; \citet{Calzetti1994}, found that \Hb/\Ha{} extinction was a factor of 2 higher than the continuum extinction, due to hot stars being associated with dusty regions. \citet{Riffel2006}, in the 0.8--2.4$\mu$m atlas of AGN, show poor correlation between the total extinction derived from \Fe{} and from \HI{} ratios, especially for starburst galaxies. This seems to apply equally for a point-by-point comparison for this object. We note that the Balmer decrement derived extinction is significantly lower than the infrared-derived values ($\sim$0.8 vs 3.4 mag); the longer IR wavelengths penetrate more of the emitting gas cloud.

\begin{figure*}[!]
	\centering
	\includegraphics[width=.75\linewidth]{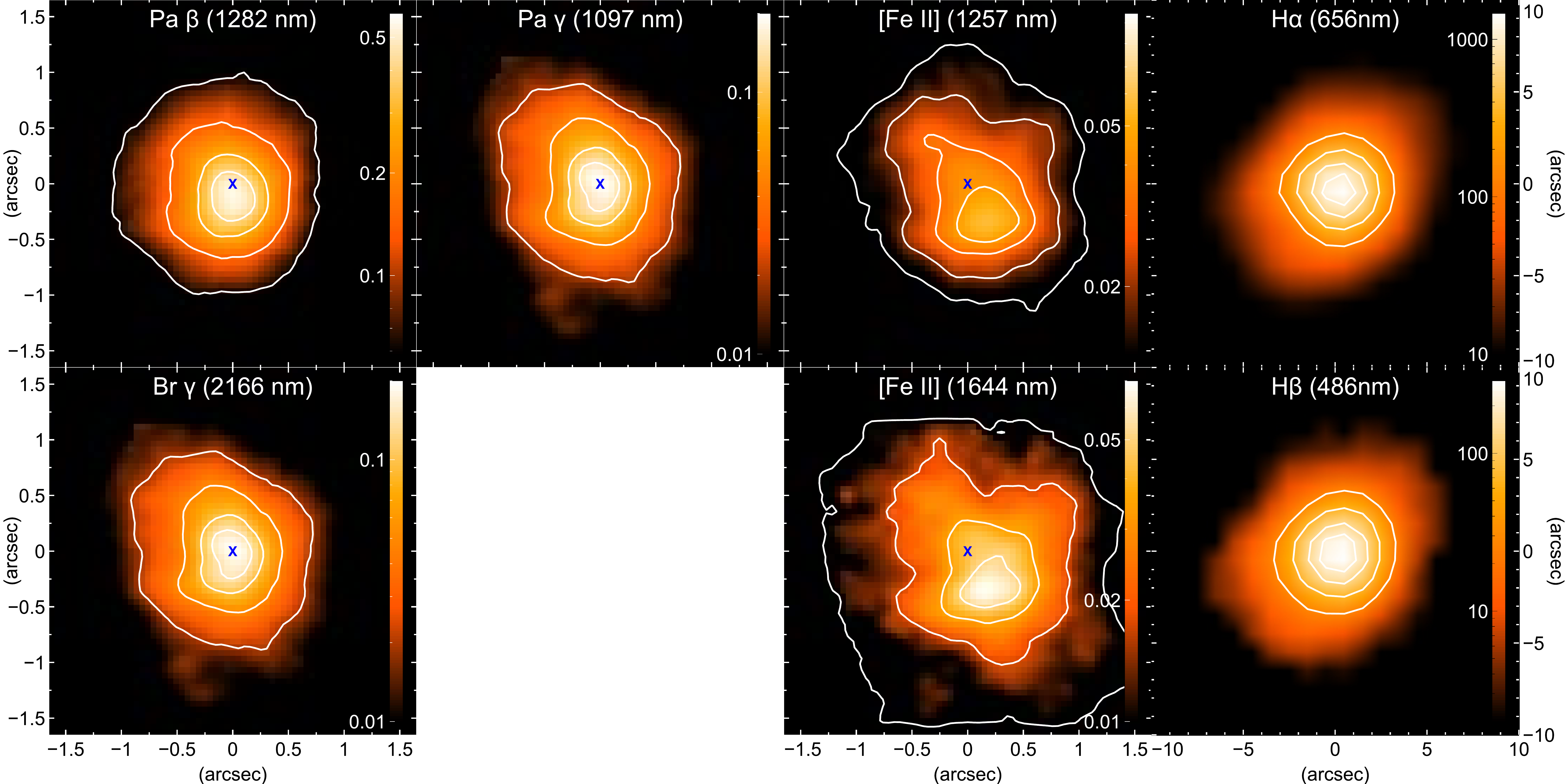}
	\caption{Emission line fluxes used for deriving extinctions. The blue cross indicates the nucleus. Flux values (shown in color bar, plotted in log scale) in units of 10\pwr{-16} \ecs. Contours are 0.1, 0.25, .5 and 0.75 of the maximum flux.  The colorbars have the maximum and minimum levels scaled so that zero extinction will have the same level for the \Fe{} and \HI{} pairs.}
	\label{fig:IC630_Fluxes_ForEC}
\end{figure*}
\begin{figure*}[!]
	\centering
	\includegraphics[width=.75\linewidth]{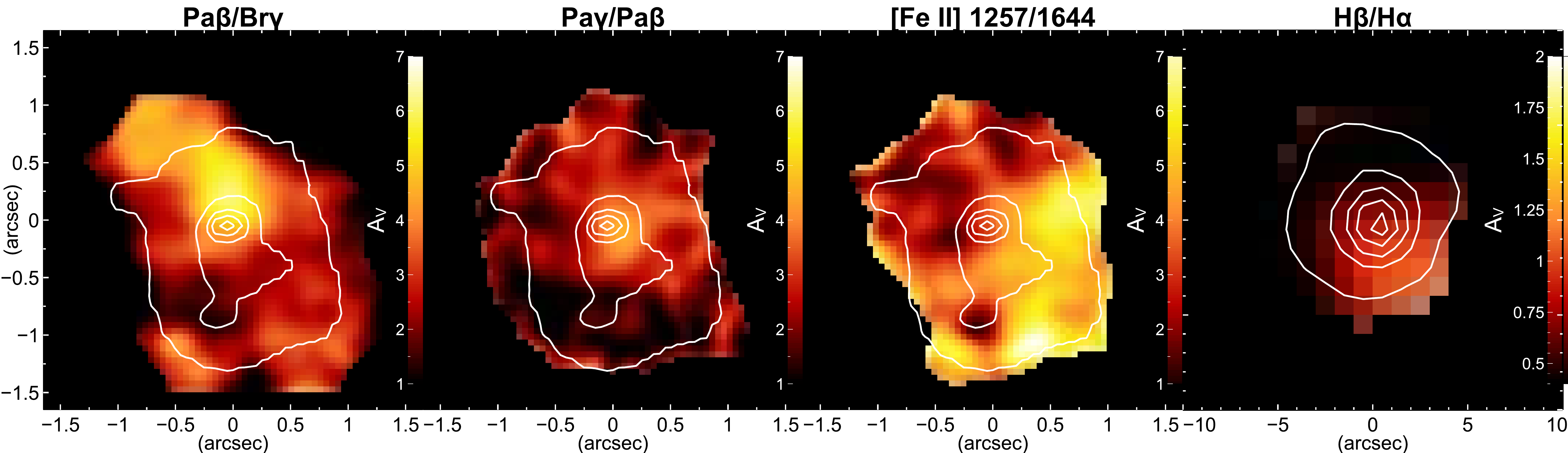}
	\caption{Extinction maps, using the emission line flux ratios as labeled. The color bars gives the A$_V$ value in magnitudes, with the infrared and visual ratios scaled to the same range. The contour plot in the IR ratio panels is the H-band flux, with the contours being 0.1--0.9 in steps of 0.2 of the maximum flux. For the visual ratio panels, the contour plot is the V-band flux with the same steps. Note the different axis scale for the V-band maps.}
	\label{fig:IC630_Extinction_Maps}
\end{figure*}
\subsection{Gas Fluxes, Kinematics, Excitation and Star Formation}
\label{sec:GasFlux}
To obtain maps of the gas emission fluxes and kinematics, we used the \textit{velmap} (velocity map) procedure in \texttt{QFitsView}. To improve S/N, especially in the velocity and dispersion measures, we applied the WVT method; the signal and noise measures are the Gaussian peak height and error, respectively. The WVT target S/N for \brg, \Fe{} and \Htwo{} was 40, 30 and 10, respectively. There were residual poor fits from noisy pixels; these were interpolated over from surrounding values on each of the derived parameters.
\subsubsection{Nebular Emission}
\begin{figure*}[!]
\centering
\includegraphics[width=1\linewidth]{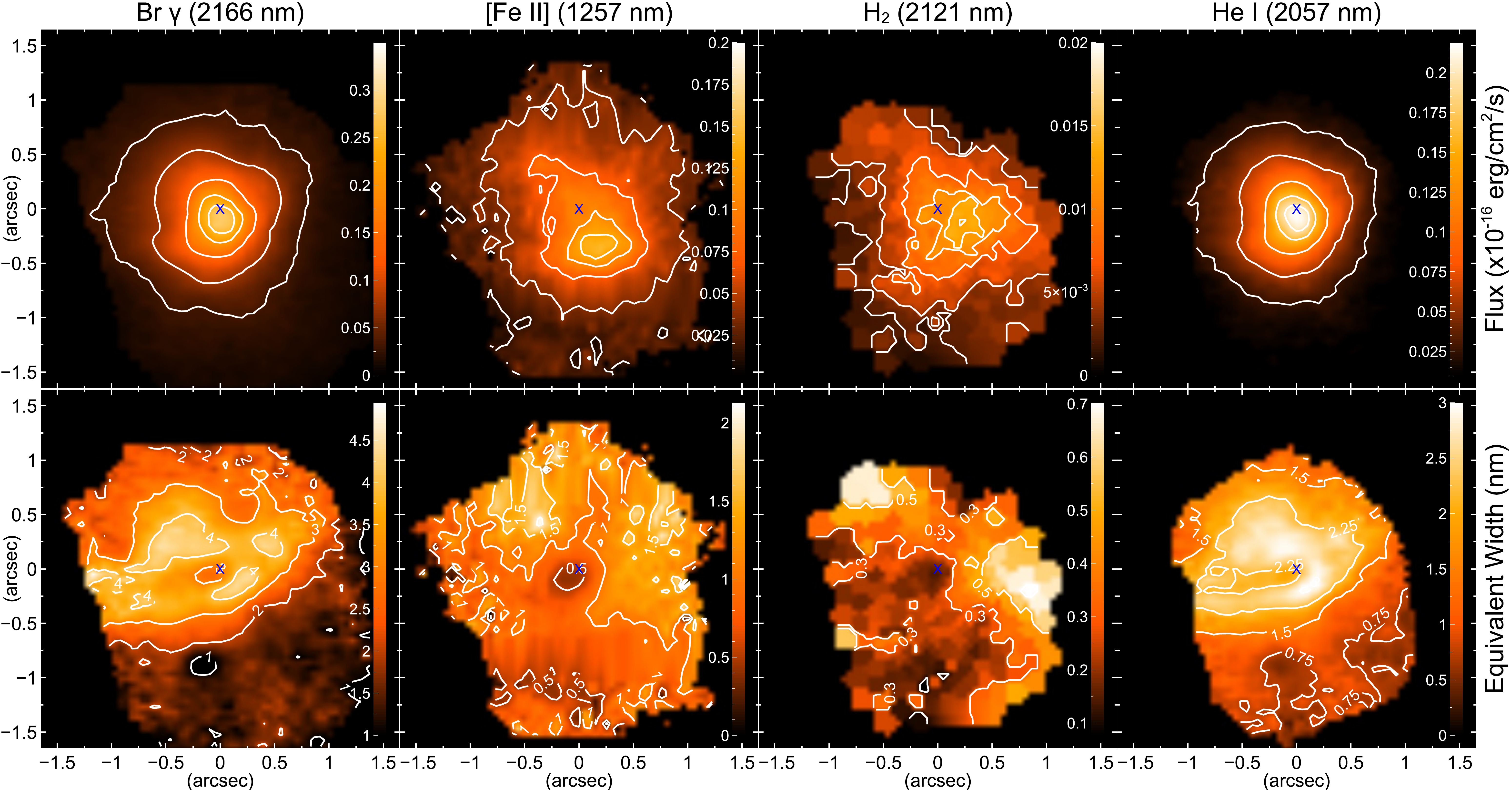}
\caption{Emission lines fluxes and equivalent widths (corrected for extinction). First row - emission line flux; scale bar is flux in units of 10\pwr{-16} \ecs, contours are at 10, 25, 50 and 75\% of the maximum value. Second row - equivalent width, scale bar in nm. The blue cross symbol marks the location of the nuclear cluster.}
\label{fig:IC630_Fluxes_EC}
\end{figure*}

Table \ref{tbl:IC630_Line_Flux} shows the relevant emission line fluxes that was measured at three locations, the nucleus and the locations of the \Fe{} and \Htwo{} maxima. The absence of high ionization species flux, indicates a lack of X-ray emission from an AGN; specifically [\ion{Ca}{8}] 2321 nm with ionization potential (IP) = 127 eV is not present. Similarly, table \ref{tbl:IC630_Line_Flux_Optical} shows the optical emission line fluxes.
\begin{table*}[!]
\centering
\caption{Integrated flux for emission lines for the nucleus (F$_1$), \Fe{} maximum (F$_2$) and \Htwo{} (F$_3$) maximum locations, with their respective uncertainties (e\_F), within $0.25 \times 0.25\arcsec$ apertures. Flux values are in 10\pwr{-16} \ecs.}
\begin{tabular}{lrrrrrrrr}
\hline
Species       & \multicolumn{1}{r}{$\lambda$ (nm) (Air)} & \multicolumn{1}{r}{$\lambda$ (nm) (Obs)} & \multicolumn{1}{r}{F$_1$} & \multicolumn{1}{r}{e\_F$_1$} & \multicolumn{1}{r}{F$_2$} & \multicolumn{1}{r}{e\_F$_2$} & \multicolumn{1}{r}{F$_3$} & \multicolumn{1}{r}{e\_F$_3$} \\ \hline
\pag          & 1094.1 & 1102.1 & 23.24 & 0.39 & 17.49 & 0.24 & 11.29 & 0.16 \\
\PII          & 1188.6 & 1197.2 &  1.45 & 0.12 &  1.03 & 0.10 & 0.744 & 0.10 \\
\Fe           & 1256.7 & 1265.8 &  3.31 & 0.15 &  3.80 & 0.11 &  3.16 & 0.07 \\
\pab          & 1282.2 & 1291.5 & 40.27 & 0.06 & 31.22 & 0.39 & 21.04 & 0.34 \\
\Fe           & 1533.5 & 1544.7 & 0.660 & 0.11 & 0.606 & 0.07 & 0.402 & 0.05 \\
\Fe           & 1643.6 & 1655.6 &  2.27 & 0.13 &  2.91 & 0.13 &  2.55 & 0.09 \\
\Htwo~1-0S(9) & 1687.7 & 1700.0 & 0.403 & 0.11 & 0.255 & 0.08 & 0.073 & 0.04 \\
\HeI          & 2059.7 & 2074.7 &  5.82 & 0.14 &  4.01 & 0.12 &  2.86 & 0.12 \\
\Htwo~2-1S(3) & 2073.5 & 2088.6 & 0.107 & 0.03 & 0.093 & 0.02 & 0.089 & 0.02 \\
\Htwo~1-0S(1) & 2121.3 & 2136.7 & 0.323 & 0.04 & 0.294 & 0.04 & 0.351 & 0.02 \\
\brg          & 2166.1 & 2181.9 &  7.57 & 0.27 &  5.60 & 0.21 &  4.01 & 0.22 \\
\Htwo~1-0S(0) & 2223.3 & 2239.5 & 0.302 & 0.02 & 0.133 & 0.02 & 0.120 & 0.02 \\
\Htwo~2-1S(1) & 2247.7 & 2264.1 & 0.118 & 0.03 & 0.093 & 0.02 & 0.118 & 0.02 \\
\Htwo~1-0Q(1) & 2406.6 & 2424.1 & 0.293 & 0.03 & 0.342 & 0.04 & 0.397 & 0.02 \\ \hline
\end{tabular}
\label{tbl:IC630_Line_Flux}
\end{table*}
\begin{table}[!]
	\centering
	\caption{Optical integrated flux for emission lines from nuclear region (2\arcsec radius). Flux values are in 10\pwr{-14} \ecs.}
	\label{tbl:IC630_Line_Flux_Optical}
	\begin{tabular}{lrrrr}
		\toprule
		Species &  $\lambda$ (Vac) &  $\lambda$ (Obs) &   Flux & eFlux \\
		        & ($\textrm{\AA}$) & ($\textrm{\AA}$) &        &  \\ \midrule
		\OII    &           3727.1 &           3754.2 &  47.30 &  2.88 \\
		\Hg     &           4341.7 &           4373.3 &  10.36 &  0.85 \\
		\Hb     &           4862.7 &           4898.1 &  27.87 &  0.78 \\
		\OIII   &           4960.3 &           4996.4 &  15.13 &  0.38 \\
		\OIII   &           5008.2 &           5044.7 &  32.19 &  0.84 \\
		\HeI    &           5877.3 &           5920.1 &   5.00 &  0.29 \\
		\OI     &           6302.0 &           6347.9 &   1.95 &  0.09 \\
		\NII    &           6549.8 &           6597.5 &  10.98 &  1.12 \\
		\Ha     &           6564.6 &           6612.4 & 164.60 &  5.38 \\
		\NII    &           6585.2 &           6633.2 &  30.62 &  0.91 \\
		\HeI    &           6680.0 &           6728.6 &   1.99 &  0.17 \\
		\SII    &           6718.3 &           6767.2 &  11.02 &  0.37 \\
		\SII    &           6732.7 &           6781.7 &  10.87 &  0.35 \\
		\SIII   &           9071.1 &           9137.1 &  21.86 &  0.59 \\ \bottomrule
	\end{tabular}
\end{table}
Fig. \ref{fig:IC630_Fluxes_EC} presents the maps of emission line fluxes and equivalent widths (EW) for the main species; each column is labeled with the species and rest frame wavelength. The EW is calculated by dividing the flux in the emission line by the height of the continuum. 
\subsubsection{Excitation}
Emission line excitation broadly falls into two categories (1) photo-ionization by a central, spectrally hard radiation field (an AGN) or by young, hot stars or (2) thermal heating, which can be either by shocks or X-ray heating of gas masses. It has been shown \citep{Larkin1998,Rodriguez-Ardila2005} that the nuclear activity for emission-line objects can be separated by a diagnostic diagram, where the log of the flux ratio of \Htwo{}($\lambda$2121nm)/\brg{} is plotted against that of \Fe($\lambda$1257nm)/\pab{}. Following the updated limits from \citet{Riffel2013a}, the diagram is divided into tree regimes; star-forming (SF) or starburst (SB) (\Htwo{}/\brg{}$<$0.4, \Fe{}/\pab{}$<$0.6), AGN (0.4$<$\Htwo{}/\brg{}$<$6, 0.6$<$\Fe{}/\pab{}$<$2) and LINER (\Htwo{}/\brg{}$>$6,\Fe{}/\pab{}$<$2). The diagnostic emission lines are convenient; the pairs are close together in wavelength, removing the dependency on calibration accuracy and differential extinction. 

\begin{figure*}[!htb]
	\centering
	\includegraphics[width=.75\linewidth]{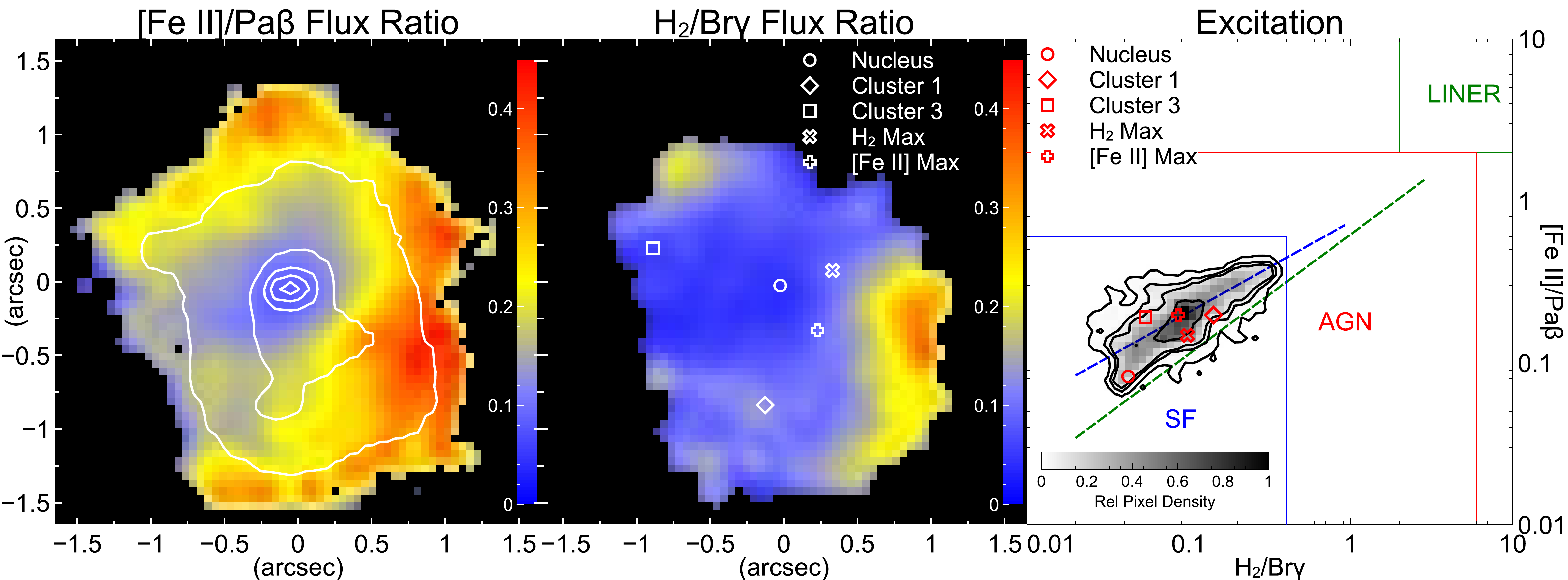}
	\caption{Excitation Diagram - Infrared. Left panel: flux ratio \Fe{}/\pab{}, with H band flux is over-plotted with white contours. Middle panel: \Htwo{}/\brg{}. Right panel: excitation map of log-log plot of flux ratios. Contour levels at 1, 5, 10 and 50\% of maximum. The locations of interest are shown with symbols in the middle and right panel. The straight-line fit to the data is plotted in blue dashed line; the fit from \citet{Riffel2013a} as a green dashed line. }
	\label{fig:IC630_Excitation}
\end{figure*}

Fig. \ref{fig:IC630_Excitation} maps the ratios for each pair of emission lines.  Specific locations of interest are also plotted with symbols; the nucleus, the \Htwo{} and \Fe{} flux maxima and clusters ``1'' and ``3'' from the continuum plots. The \Fe/\pab{} ratio has a range of 0.03 to 0.44, with the lowest value in the center and the highest value in the SW region, located about 215 pc from the center. The \Htwo{}/\brg{} ratio has a range 0.03 to 0.37, with a similarly located peak to the SW.

Fig. \ref{fig:IC630_Excitation} also plots the density diagram for the values at each spaxel of log(\Fe/\pab) against log(\Htwo/\brg), with the excitation mode regions (SF, AGN and LINER) delineated and with the locations of interest plotted with symbols. The straight-line fit to the points (blue line in Fig. \ref{fig:IC630_Excitation}) is:
\begin{multline}
	\log\left(\text{[Fe \sc{ii}]}/\text{Pa}\beta\right)= \\
	0.558(\pm0.1)\times\log\left(\text{H}_2/\text{Br}\gamma\right)-0.130(\pm0.091)
\end{multline}
The high correlation coefficient ($r=0.79$) indicates that this relationship can be used to determine the excitation mode from just one set of measurements, e.g. from the J band spectrum when K band is not available.

\begin{figure*}[!htb]
	\centering
	\includegraphics[width=.75\linewidth]{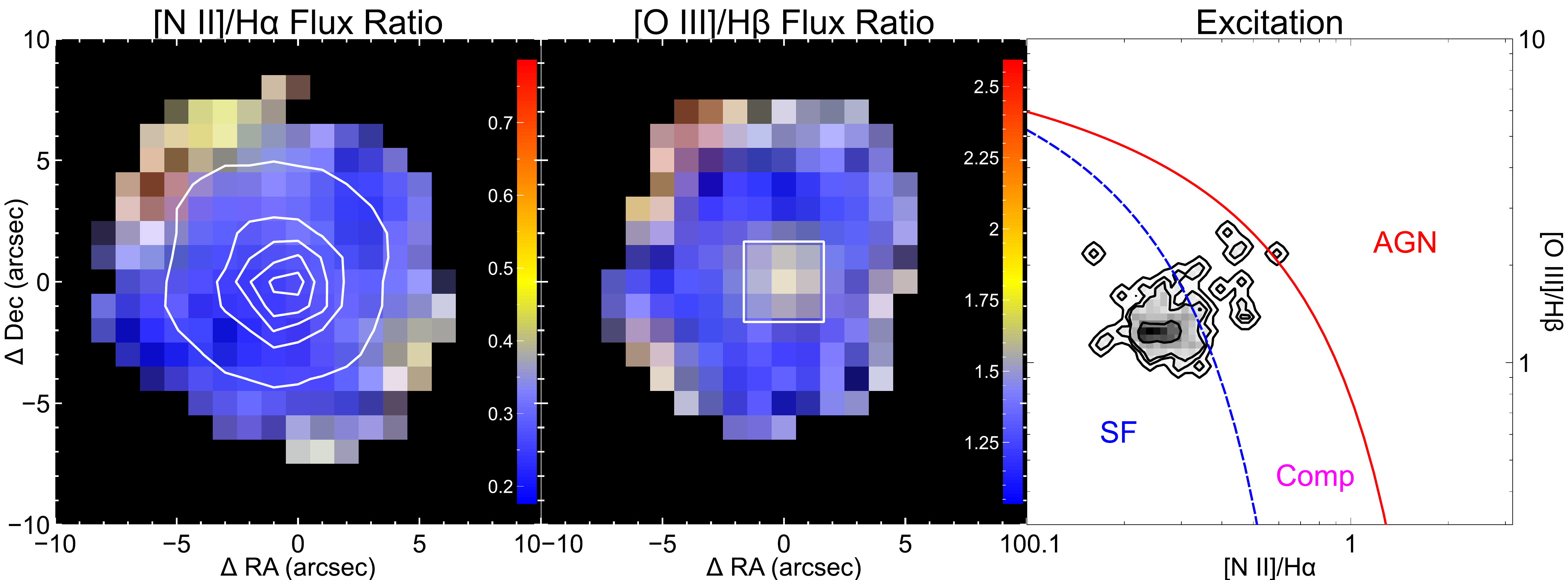}
	\caption{Excitation Diagram - Optical. Left panel: flux ratio \NII/\Ha{} (V-band flux overplotted in contours of 10,30,50,70 and 90\% of maximum). Middle panel: \OIII/\Hb (white square is FOV of the IR IFU instruments, for comparison). Right panel: excitation map of log-log plot of flux ratios. Contour levels at 1, 5, 10 and 50\% of maximum.}
	\label{fig:IC630_Excitation_Optical}
\end{figure*}

Similarly, Fig. \ref{fig:IC630_Excitation_Optical} plots the BPT excitation diagram and the spaxel density diagram for the optical, as defined in \citet{Kewley2006}, using the \NII/\Ha~ and \OIII/\Hb~flux ratios.  

\subsubsection{Gas Kinematics}
\begin{figure*}
	\centering
	\includegraphics[width=1\linewidth]{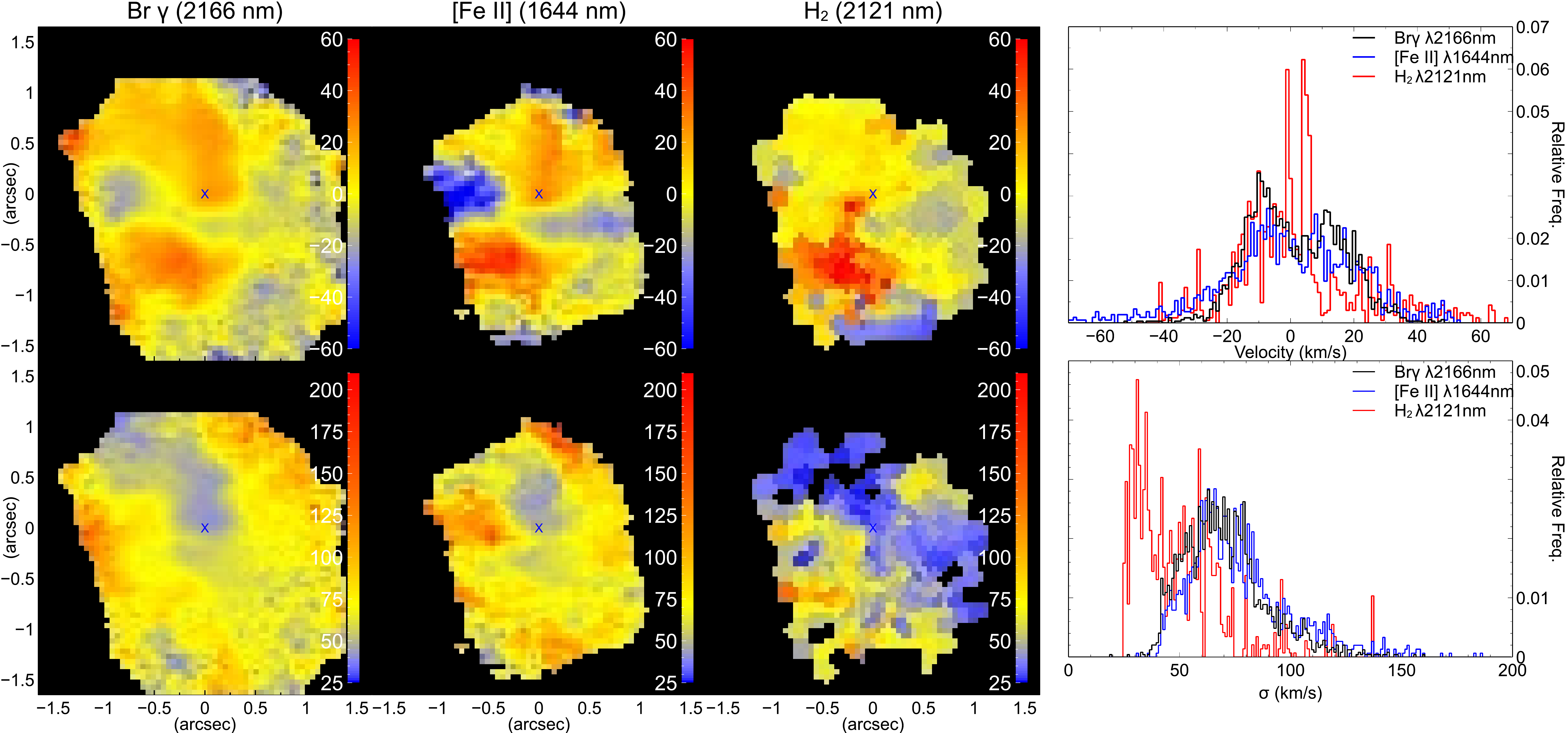}
	\caption{LOS velocity and dispersion for \brg{}, \Fel1656nm and \Htwo. Top row - velocity. Bottom row - dispersion. Right panels: velocity and dispersion histograms. All values are in \kms. }
	\label{fig:IC630_Vel_Sigma}
\end{figure*}

\begin{figure}
	\centering
	\includegraphics[width=1\linewidth]{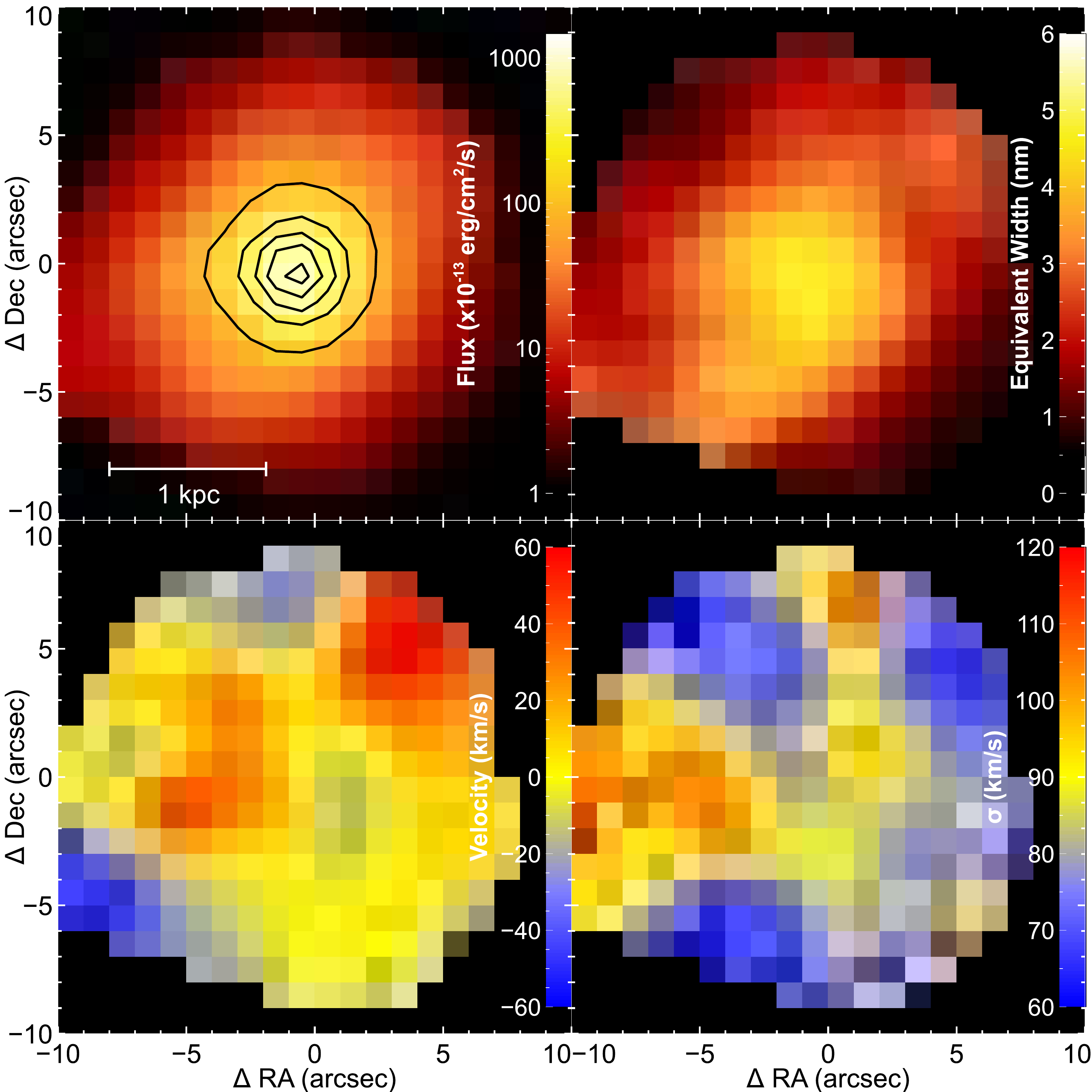}
	\caption{\Ha{} flux and kinematic maps. Top left: flux (units of 10\pwr{-13} \ecs) (V-band flux overplotted in contours of 10,30,50,70 and 90\% of maximum). Top right: equivalent width in nm. Bottom left: LOS velocity. Bottom right: LOS dispersion.}
	\label{fig:IC630_Halpha_Velmap}
\end{figure}

Fig. \ref{fig:IC630_Vel_Sigma} presents maps of the line-of-sight (LOS) velocities and dispersions of the main  species; \brg{}($\lambda$~2166nm), \Fe{}($\lambda$~1644nm) and \Htwo{}($\lambda$~2121.5nm). All LOS velocities have been set so that the zero is the median value. We also present the maps for \Ha{} (Fig. \ref{fig:IC630_Halpha_Velmap}), showing the flux, equivalent width, LOS velocity and dispersion. We compared the systemic velocity fields of the stars and gas from the central wavelength average around the nucleus, after reduction to rest frame; these vary from +7 \kms{} (\brg) to +40 \kms{} (stars). These are all within the measurement error ($ \pm $55 \kms{} for NIFS and $ \pm $125 \kms{} for SINFONI). Heliocentric corrections for different observations dates were $<$7\kms.

\subsubsection{Channel Maps}
\begin{figure*}[!htb]
	\centering
	\includegraphics[width=.75\linewidth]{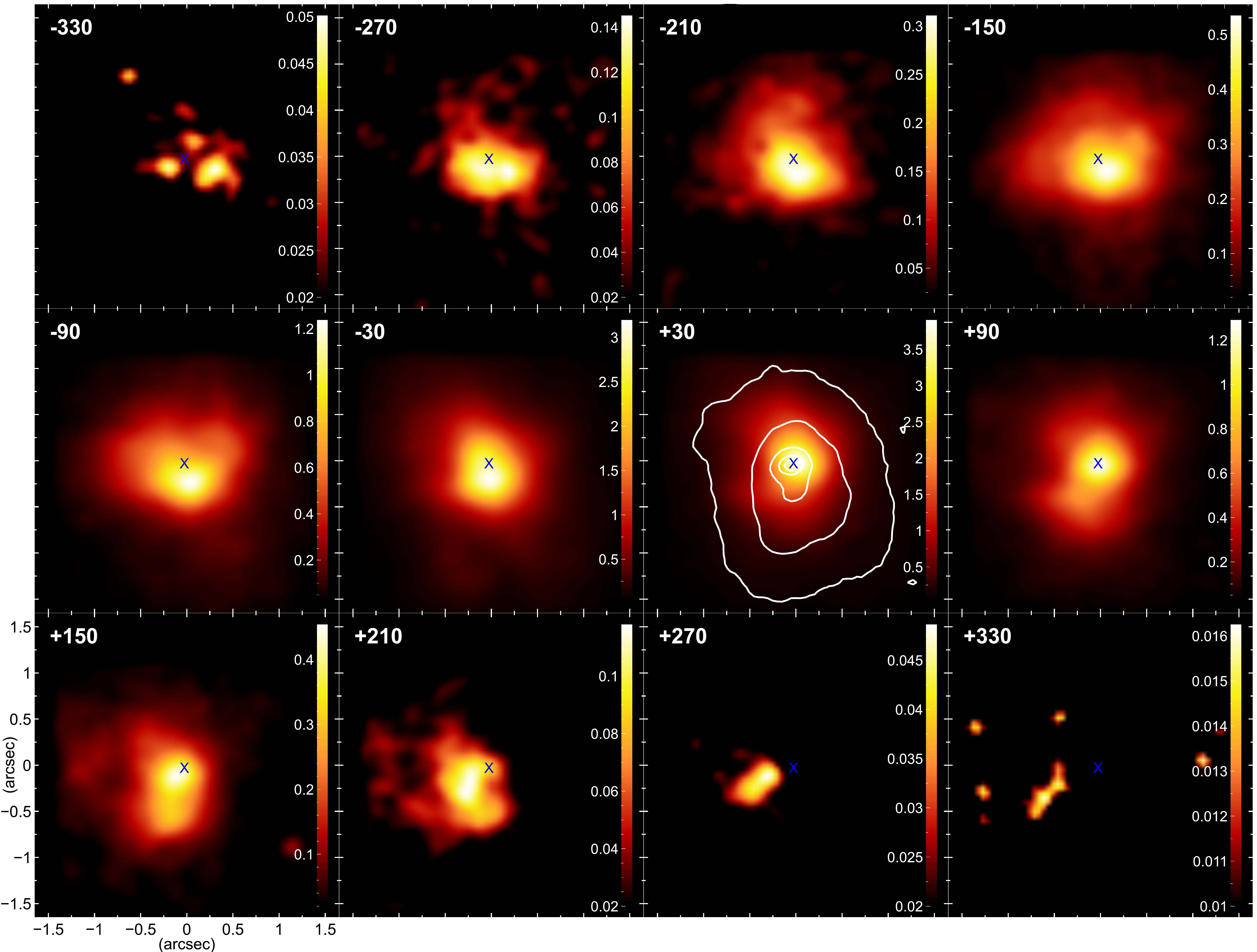}
	\caption{Channel map for \brg. Each channel is has a width of 60 \kms, with the channel central velocity shown in the top left corner of each plot. The blue cross is the position of the nucleus. Color values are fluxes in units of 10\pwr{-18} \ecs. Positive velocities are receding, negative are approaching. The white contour on the +30 \kms{} channel is the K band continuum flux, at levels of 10, 25, 50, 75 and 90 \% of maximum.}
	\label{fig:IC630_CHMAP_2182}
\end{figure*}

\begin{figure*}[!htb]
	\centering
	\includegraphics[width=.55\linewidth]{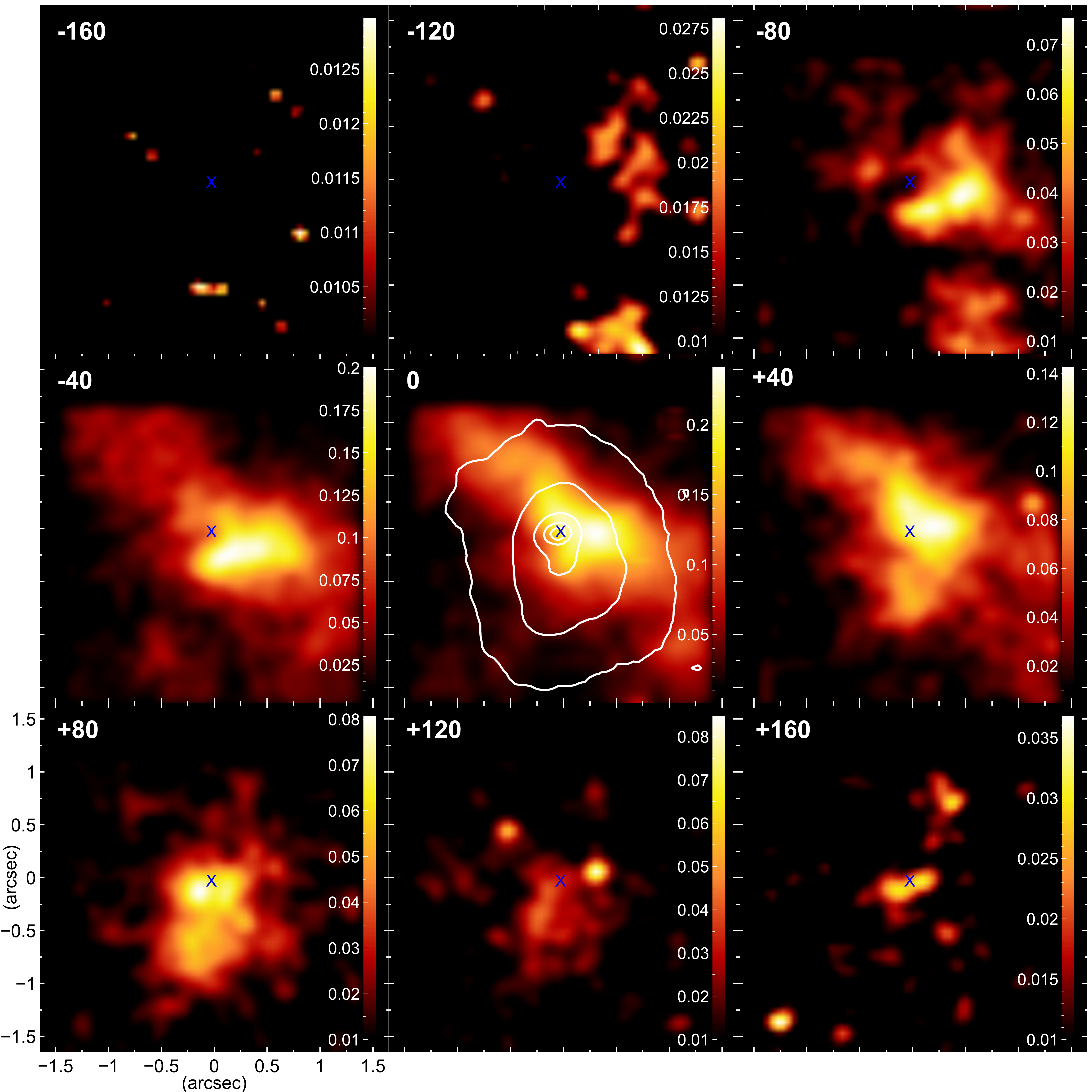}
	\caption{Channel map for \Htwo. Channel widths are 40 \kms. Values are fluxes in units of 10\pwr{-18} \ecs. Contours as per Fig. \ref{fig:IC630_CHMAP_2182}.}
	\label{fig:IC630_CHMAP_2137}
\end{figure*}

\begin{figure*}[th]
	\centering
	\includegraphics[width=.75\linewidth]{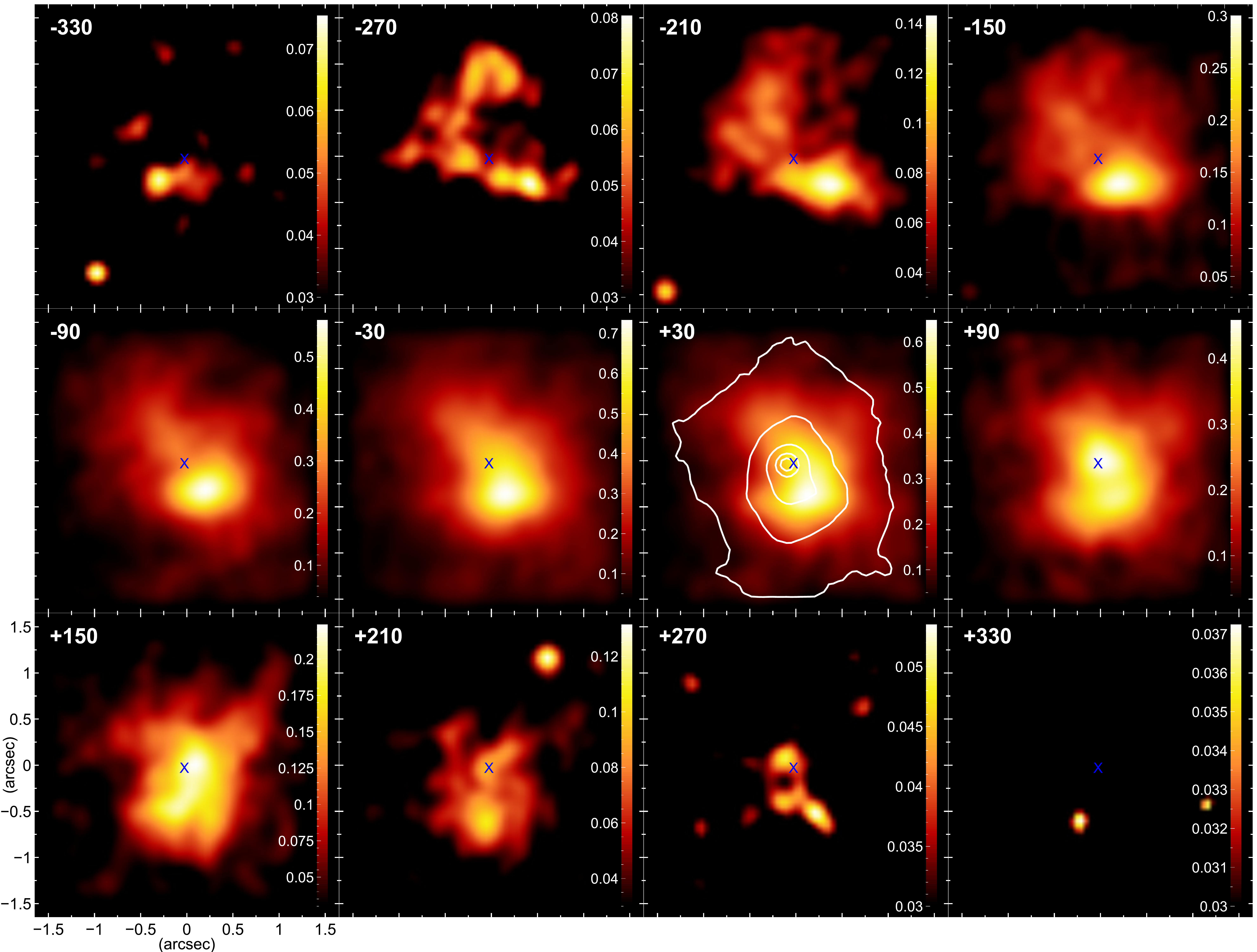}
	\caption{Channel map for \Fel1644 nm. Channel width are 60 \kms. Values are fluxes in units of 10\pwr{-18} \ecs. Contour is the J band continuum flux, as in \ref{fig:IC630_CHMAP_2182}}
	\label{fig:IC630_CHMAP_1266}
\end{figure*}

In order to map the flux distributions at all velocities covered by the emission-line profiles and to assist delineating gas flows, we construct channel maps along the profile of the  \brg{}, \Fel1644 nm and \Htwo{} emission. The maps were constructed by subtracting the continuum height, derived from the velocity map function \texttt{velmap}, from the data cube. The spectral pixels are velocity binned and smoothed to reduce noise. Figs. \ref{fig:IC630_CHMAP_2182}, \ref{fig:IC630_CHMAP_2137} and \ref{fig:IC630_CHMAP_1266} show the derived channel maps for \brg{}, \Htwo{} and \Fel1644 nm, respectively. The \Fel1644nm line is used rather than the \Fel1257 nm line, because of the J cube instrumental fingerprint. Note that the fluxes are rescaled on each map to bring out the structure, rather than share a common scale across all maps. 
\section{Discussion}
\subsection{Black Hole Mass and Stellar Kinematics}
The stellar kinematic results presented above show no ordered rotation and have a flat velocity dispersion structure. We can therefore conclude that in the central region, the stellar dynamics are either face-on or pressure supported. The first is favored from the associated gas dynamics (see below), and the fact that this object is a S0 galaxy, which will have stellar rotation.

From \citet{Ho2008}, the X-ray luminosity indicates a spectral class somewhere between Seyfert 1 and 2. The measured $R_{Edd}$ is somewhat high for this class range, but this could be due to the uncertainties in the \mbh{} measurement, plus emission from SN remnants and unresolved point sources, e.g high-mass X-ray binaries \citep{Mineo2013}. This could be the major component of the X-ray emission, with minimal contribution from any SMBH. We note also that the uncertainties in $R_{Edd}$ are not inconsistent with a value of zero, i.e. with no SMBH luminosity and X-ray emission solely from star formation.

\subsection{Emission Line Properties and Diagnostics}
\label{sec:elprops}
The \brg{} and \HeI{} fluxes show strong central concentration, contiguous with the nuclear cluster. This emission is from star-forming regions. On the other hand, the \Fe{} and \Htwo{} fluxes are located distinctly off-center. The \Fe{} peak is some 55 pc to the SW of the central cluster, roughly contiguous with the ridge feature ``2''. The \Htwo{} flux maximum is at a  different location due west about 50 pc from the center. 
 
The \brg{} and \HeI{} equivalent widths show a similar structure; there is relatively less star-formation in the nuclear cluster, with the peak being in a rough ring about 50--80 pc from the center. The \Fe{} and \Htwo{} maximum values  track their respective flux structures; the central ``hole'' is also visible in these maps.

The \Fe{} and \PII$\lambda$~1188.6 nm emission lines can be used to diagnose the relative contribution of photo-ionization and shocks \citep{Storchi-Bergmann2009}, where ratios $\sim$ 2 indicate photo-ionization (as the \Fe{} is locked into dust grains), with higher values indicating shocked release of the \Fe{} from the grains (up to 20 for SNRs). The flux ratio over the field varies from 2.7 in the nucleus to 7.4 at the highest excitation ratio \Fe/\pab{} location (in the SW corner of the field); this indicates that photo-ionization is the main excitation mode over most of the field, with an increased shock contribution at the more AGN-like excitation locations.

To determine the electron density, we use the ratio of \Fe$\lambda~1533/\lambda$~1644 and the method of \citet{Storchi-Bergmann2009}. Using the diagrams in that paper's Fig. 11, we measured this ratio over a region of $0.25\times0.25\arcsec$ at the nucleus ($0.27\pm0.05$) and at the location of the maximum \Fe{} emission ($0.19\pm0.02$), deriving  values of $\sim$ 32000 cm\pwr{-3} (nucleus) and 8000 cm\pwr{-3} (\Fe{} maximum). Over the whole field, the \Fel1533 nm flux was difficult to determine (having considerable noise), but values in the range $14000 < n_{e} < 60000$ cm\pwr{-3} are obtained where it can be measured.  These values are consistent with their findings for NGC~4151. We can also use the ratio \Fe$\lambda~1533/\lambda$~1257 \citep{Nussbaumer1988}, and obtain similar values, $10000 < n_{e} < 32000$ cm\pwr{-3}.

To determine if \Htwo{} excitation results from soft-UV photons (from star formation) or thermal processes (from shocks or X-ray heating), we use the line ratio \Htwo{}$\lambda2247.7/\lambda2121.8$ \citep{Riffel2006a}. This has a value of $\sim0.1-0.2$ for thermal and $\sim0.55$ for fluorescent processes. At the location of the maximum \Htwo{} flux, the ratio is measured as $0.21\pm0.06$, indicating thermal processes. Since AGN-like excitation is minimal over the whole field and therefore X-rays from an accretion disk heating are absent, we conclude that shocks are the main excitation mode for \Htwo{}. The line \Htwo{} 2–-1 S(3) ($\lambda$~2073.5 nm) can also be used to investigate the \Htwo{} excitation. In the case of X-ray irradiated gas, this line is expected to be absent \citep{Davies2005}. This line is present, supporting the assertion that shocks are the main excitation mechanism. Other \Htwo{} lines ($\lambda$~1957 and $\lambda$~2033 nm) to place on the \citet{Mouri1994} diagrams were not present or outside the spectral range.

To determine the thermal excitation temperature for \Htwo{}, we use the observed fluxes at the available emission. Following \citet{Wilman2005} and \citet{Storchi-Bergmann2009}, the relationship is:
\begin{equation}
\log\left( \dfrac{F_{i}~\lambda_{i}}{A_{i}~g_{i}}\right)  = constant - \frac{T_{i}}{T_{exc}}
\end{equation}
where $F_i$ is the flux of the \textit{i}th \Htwo{} line, $\lambda_i$ is its wavelength, $A_i$ is the spontaneous emission coefficient, $g_i$ is the statistical weight of the upper level of the transition, $T_i$ is the energy of the level expressed as a temperature and $T_{exc}$ is the excitation temperature. This relation is valid for thermal excitation, under the assumption of an \textit{ortho:para} abundance ratio of 3:1. The \textit{A, g} and $T_{exc}$ for each line was obtained from on-line data ``Molecular Hydrogen Transition Data''\footnote{\url{www.astronomy.ohio-state.edu/~depoy/research/observing/molhyd.htm}}.

\begin{figure}[!htb]
	\centering
	\includegraphics[width=.75\linewidth]{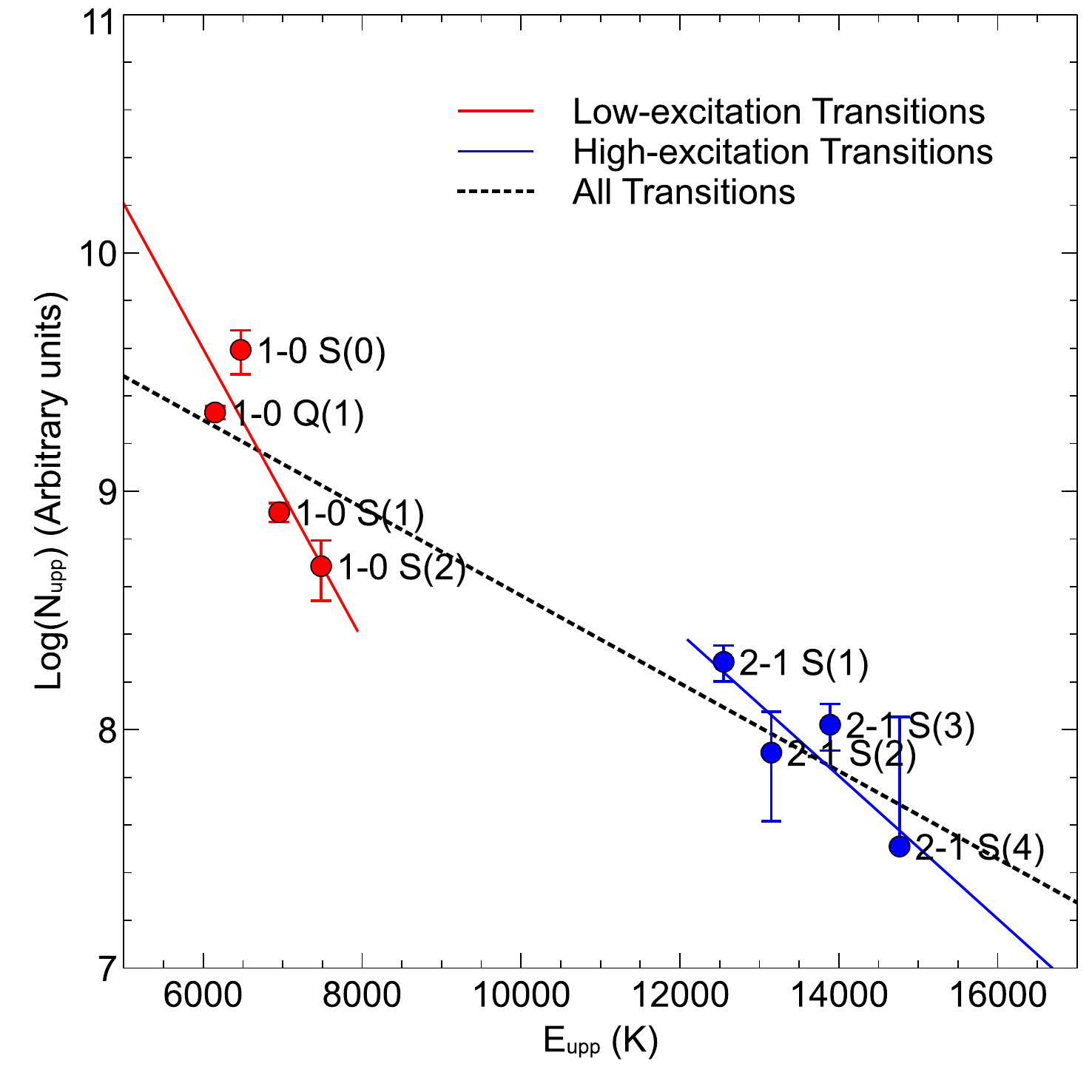}
	\caption{Relationship between log(N$_{upper} $) and E$_{upper} $ for the \Htwo{} transitions. The fits are to the low-excitation (red) and high-excitation (blue) transitions, as well as for all transitions (black). The individual transitions are labeled.}
	\label{fig:IC630_H_2_Temperatures}
\end{figure}

To improve the signal to noise, the spaxels in the K-band data cube were summed where the \Htwo{} flux was greater than $0.5 \times 10$\pwr{-18} \ecs, and the fluxes of the resulting spectrum were measured by fitting a Lorentzian curve to each spectral line. As seen in Fig. \ref{fig:IC630_H_2_Temperatures}, the observed fluxes fit the equation well (as shown by the fitted line). The resulting excitation temperature (the inverse of the fitted line slope) are $T_{exc}$ = 6135 K. This is much higher than found for several Seyfert galaxies (e.g. \citet{Riffel2015,Storchi-Bergmann2009,Riffel2014a,Riffel2011,Riffel2010a}) which are in the range 2100 -- 2700 K and in fact is above the \Htwo{} dissociation temperature of $\sim4000$ K. However, following \citet{Wilman2005}, we can postulate a two temperature model, where the lower-excitation transitions are thermalised, and the higher-excitation transitions come from either hotter gas close to the dissociation temperature, non-thermal fluorescent emission in low density gas or excitation by secondary electrons in the cloud. If we separately fit to the lower- and higher-excitation lines, we derive temperatures of $1640\pm470$ K and $3330\pm880$ K, respectively. We can also compute the ro-vibrational temperatures from the ratios \Htwo{} 1-0 S(0)/\Htwo 1-0 S(2) and \Htwo{} 2-1 S(1)/\Htwo{} 1-0 S(1) \citep{Busch2016}, and derive T$_{rot(\nu=1)}=530$ K and T$_{vib}=3870$ K.

We can also calculate the electron temperature and density from the optical spectrum, using the method of \citet{Kaler1986} for the temperature from the \NII{} $I(\lambda6548+\lambda6584)/I(\lambda5754)$ ratio and the method of \citet{Acker1986} for the density from by the \SII{} $I(\lambda6717)/I(\lambda6731)$ ratio. In the inner 2\arcsec, the temperature is $\sim$ 8630 K and the density is $\sim$ 670 cm\pwr{-3}. In the annulus from 2 to 5\arcsec, the temperature and density are calculations are more uncertain, due to the weakness of the \NII $\lambda5754$ line; they are 7500$ \pm $2500 K and 280$\pm$40 cm\pwr{-3}.
\clearpage

\subsection{Gas Masses}
We can derive the gas column density from the visual extinction value. The gas-to-extinction ratio, $N_{H}/A_{V}$, varies from 1.8 \citep{Predehl1995} to 2.2 $\times$ 10\pwr{21} cm\pwr{-2} \citep{Ryter1996}; we will use a value of 2.0 $\times$ 10\pwr{21} cm\pwr{-2}. We thus derive the relationship:
\begin{equation}
\sigma_{Gas} = 22.1~A_V~\mmsun{}~pc^{-2} \label{eqn:ISM}
\end{equation}
Using the extinction map from the \pag/\pab{} ratio, we derive the total gas mass in the central region, as given in table \ref{tbl:GasMass} below.
We also derive the ionized hydrogen and warm/hot and cold \Htwo{} gas masses, using the formulae from \citet{Riffel2013b,Riffel2015}:
\begin{eqnarray}
M_{HII} &\approx& 3 \times 10^{17} F_{BR\gamma}d^2~\mmsun  \label{eqn:HII}\\
M_{H_2} (hot) &\approx& 5.0776 \times 10^{13} F_{H_2}d^2~\mmsun  \label{eqn:H2H}\\
M_{H_2} (cold) &\approx& 1174 \dfrac{L_{H_{2}}}{\mlsun}~\mmsun \nonumber\\
&\approx& 3.65 \times 10^{19} F_{H_2}d^2~\mmsun  \label{eqn:H2C}
\end{eqnarray}
where \textit{d} is the distance in Mpc and \textit{F} is measured in erg cm\pwr{-2} s\pwr{-1}. The \Htwo{} flux is that of the 2121 nm line. The cold-to-warm molecular gas mass ratio (7.2 $\times$ 10\pwr{5}) is originally derived in \citet{Mazzalay2013a} from the observed CO radio emission with estimates of CO/\Htwo{} ratios (which can vary over a range 10\pwr{5} to 10\pwr{7}); the figures derived in table \ref{tbl:GasMass} for the cold \Htwo{} are therefore only an estimate.
\begin{table*}[!htb]
	\centering
	\caption{Gas Masses. Masses within the maximum value pixel, within 100 and 200 pc of center and over the whole field, plus the surface density within the central 100 and 200 pc. Values are derived from equations \ref{eqn:ISM} - \ref{eqn:H2C}.}
	\footnotesize
	\begin{tabular}{lcccccc}
		\hline
		Species              & Max. Pixel\tablenotemark{a} \msun pc\pwr{-2} &     \msun (100pc)      &     \msun (200pc)      & \msun (Total)\tablenotemark{b} & \msun pc\pwr{-2} (100 pc) & \msun pc\pwr{-2} (200 pc) \\ \hline
		ISM ($\sigma_{Gas}$) &                     101                      & 2.0 $\times$ 10\pwr{6} & 6.1 $\times$ 10\pwr{6} &               --             &           63.7            &           48.8            \\
		\HII                 &                     8.5                      & 2.0 $\times$ 10\pwr{6} & 3.4 $\times$ 10\pwr{6} &     3.7 $\times$ 10\pwr{6}     &           63.7            &           27.1            \\
		\Htwo{} (warm)       &           7.1 $\times$ 10\pwr{-5}            &          23.3          &           46           &               49               &  7.4 $\times$ 10\pwr{-4}  &  3.7 $\times$ 10\pwr{-4}  \\
		\Htwo{} (cold)       &                     51.1                     & 1.7 $\times$ 10\pwr{7} & 3.3 $\times$ 10\pwr{7} &     3.5 $\times$ 10\pwr{7}     &            541            &            264            \\ \hline
	\end{tabular}
	\tablenotetext{a}{{\footnotesize Pixel with greatest flux or highest density}}
	\tablenotetext{b}{{\footnotesize Over whole field where valid measurement}}
	\label{tbl:GasMass}
\end{table*}

The cold \Htwo{} mass estimate is greater than the ISM mass as derived from extinction, even at the lower estimate of the scaling relationship with warm \Htwo{}. This could be because in this environment the dust grains which cause the extinction are being evaporated by the star formation photo-ionization, thus reducing the gas-to-dust/extinction ratio, and underestimating the ISM mass. 

\cite{Schonell2017} summarizes results for 5 Seyfert 1, 4 Seyfert 2 and 1 LINER galaxies observed by the AGNIFS group; the range of \HII{} surface densities is 1.5--125 \msun/pc\pwr{2} and of \Htwo{} (cold+warm) surface densities is 526--9600 \msun/pc\pwr{2}. Our values (64 and 540 \msun/pc\pwr{2}, respectively) are within these ranges, with the \Htwo{} value on the lower end of the range.
\subsection{Star Formation and Supernovae}
In this object, SF and SNe supply the bulk of non-stellar emission, with minimal AGN activity. We can derive the SF rates, using relationship between star formation rate and hydrogen recombination lines from \citet{Kennicutt2009}, which is regarded as an ``instantaneous'' trace of SF. Assuming Case B recombination at T$_{e}$ = 10\pwr{4}K and applying the line-strength ratios, we get:
\begin{equation}
SFR(\mmsun yr^{-1}) = 5.66 \times 10^{-40} L(Br\gamma)~[erg~s^{-1}]
\end{equation}
The results are shown in table \ref{tbl:SFRSNR}.

The star formation rates given in table \ref{tbl:SFRs} above are mostly within a factor of 2, which is also in good agreement with the \brg{} derived rate as in table \ref{tbl:SFRSNR}. The exceptions are the WISE W4 and the X-ray derived values. The WISE color-color diagram \citep{Wright2010} places this object in the starburst/LIRG region. The X-ray flux from the nucleus could include AGN activity, which would reduce the derived SFR. There may be substantial highly obscured star formation which would only manifest in far IR and Xrays, which penetrate the dust. From the SED fitting, the derived SFR is about $\sim1.1~\mmsun~yr^{-1}$. The SFR derived from the radio flux is in excellent agreement with the other indicators, showing there is no AGN component emission. The origin of radio emission from radio-quiet quasars has been debated; \citet{HerreraRuiz2016} finds that this can come from the AGN. This object provides a counter-example. 

\citet{Rosenberg2012} presents a formula for the supernova (SN) rate, based on the measured \Fel1257nm flux. SN remnant (SNR) shock fronts destroy dust grains by thermal sputtering, which releases the iron into the gas phase where it is singly ionized by the interstellar radiation field. In the extended post-shock region, \Fe{} is excited by electron collisions \citep{Mouri2000}, making it a strong diagnostic line for tracing shocks. Their relationship is
\begin{eqnarray}
	&log (\nu_{SNR}~[yr^{-1}~pc^{-2}])  = \nonumber\\&(1.01 \pm 0.2)~\log (L_{[Fe II]1257}~[erg~s^{-1}~pc^{-2}]) \nonumber\\&- 41.17 \pm 0.9
\end{eqnarray}
The results, using the de-reddened \Fe{} flux, are shown in table \ref{tbl:SFRSNR}.

The SN rate derived from radio emission using the indicator from \citet{Condon1992} can be compared with that derived from the \Fe{} flux, above. For the VLSS (1.4 GHz) and TGSS (150 MHz) fluxes, $\nu_{SNR}$ = 0.089 and 0.035 yr\pwr{-1}, respectively. These are higher than the value from the \Fe{} flux of 0.0083 yr\pwr{-1}, however, the radio values are for the whole galaxy, not just the inner 500 pc.
\begin{table*}[!]
	\centering
	\caption{Star formation rate and supernova rate. Regions as for table \ref{tbl:GasMass}. The SNR is shown as the rate per year per pc\pwr{2}, per year over the region and the interval between SN.}
	\footnotesize
	\label{tbl:SFRSNR}
	\begin{tabular}{lcccc}
		\toprule
		                            &        Max. Pixel         &         100 pc          &         200 pc          &          Total          \\ \midrule
		SFR (\msun yr\pwr{-1})      & 2.1 $\times$ 10\pwr{-3} &          0.44           &          0.77           &           0.83           \\
		SNR (yr\pwr{-1} pc\pwr{-2}) & 1.9 $\times$ 10\pwr{-7}  & 1.1 $\times$ 10\pwr{-7} & 5.8 $\times$ 10\pwr{-8} & 2.9 $\times$ 10\pwr{-8} \\
		SNR (yr\pwr{-1})            & 1.3 $\times$ 10\pwr{-5}  & 3.4 $\times$ 10\pwr{-3} & 7.3 $\times$ 10\pwr{-3} & 8.5 $\times$ 10\pwr{-3} \\
		SN Interval (yr)            &          76500           &           300           &           135           &           118            \\ \bottomrule
	\end{tabular}
\end{table*}

The star formation rate for the whole field given above can be compared with the rate derived from the WiFeS data. The \Ha{} flux from the nuclear $3\times3\arcsec$ is 1.5$\times10^{-12}$ \ecs, giving an emitted flux of 2$\times10^{-39}$~erg~s\pwr{-1}. The SFR is determined by \citep{Kennicutt2009}:
\begin{equation}
SFR(\mmsun~yr^{-1} ) = 5.5~\times 10^{-42}~L_{\textrm{H}\alpha}~[erg~s^{-1}]
\end{equation}
giving 1.1 \msun~yr\pwr{-1}, in excellent agreement with the IR-derived value.
\subsection{Stellar Population}
Correcting the \textit{H--K} color map for the average visual extinction of 3.4 mag decreases the values by 0.2 mag, 95\% of the pixels are in the range 0.37 to 0.58. Given the uncertainties about the absolute flux calibration of the data cubes, we are cautious about assigning a stellar type to the colors (e.g. the tables of \citep{Bessell1988}); however the relative differences are clear, with the clusters about 0.3--0.4 mag bluer than the rest of the field. The average color over the central nucleus is 0.26, the color of an M star. \citet{Persson1983}, in a study of clusters in the Magellanic clouds, showed that an admixture of luminous, intermediate age ($1-8\times10^9$ Myr) carbon stars can cause very red \textit{H--K} colours.

The mass of the nuclear cluster can be estimated from the cluster size and velocity dispersion; using the virial formula:
\begin{equation}
M \approx \frac{3}{2}~\dfrac{\sigma_{R}^{2}~R}{G}
\end{equation}
where $R$ is the cluster radius and $\sigma_{R}^{2}$ is the dispersion, and using the values of 50 pc and 44 \kms{}, we derive a mass of $\sim 3.4 \times 10^{7} \mmsun$. This is a lower limit; any rotation will support more mass and the stars are probably not settled into  virial equilibrium. Simulations (see section \ref{sec:SimulCompare}) suggest that the stars settle into a thick disk, rather than a spherical cluster. Using the luminosity of 9.0 $\times$ 10\pwr{7} \lsun{}, calculated above, we get a mass to light ratio of $\sim$ 0.38. This is again consistent with an old stellar population (with a M/L of about 1) mixed with younger star formation.

In \citet{Balcells2007}, the sizes of nuclear disks are measured by analyzing surface brightness profiles of S0--Sbc galaxies from Hubble Space Telescope NICMOS images. They find that the central star clusters are usually unresolvable at $\sim$ 10 pc resolution, whereas the nuclear disks sizes are found to be in the range 25--65 pc (from their table 4). It can be deduced that the central object visible in our observations is more likely to be a disk than a spherical structure.

The ratio of \HII{} to \HeI{} emission can be used as an indicator of \textit{relative} age of star clusters, following \citet{Boker2008}; as the \HeI{} ionization energy is 24.6eV vs. 13.6eV for hydrogen, the \HeI{} emission will arise in the vicinity of the hotter stars (B- and O-type), which will vanish fastest. Taking the ratio of the flux from \brg{} and \HeI$\lambda$~2058nm for the nuclear cluster and the two light concentrations and the ridge extension, the nuclear cluster shows a value of 0.72, while the other three show value of 0.60--0.63. Even though the differences are small, it is an indicator that the nuclear cluster is the youngest region.

We can estimate the stellar ages in the central region, using the methods of \citet{Brandl2012} and references therein, which examines the starburst ring in NGC~7552. They use the pan-spectral energy distribution of starburst models of \citet{Groves2007} to derive cluster ages based on \brg{} EW. The map for \brg{} (in Fig. \ref{fig:IC630_Fluxes_EC}) shows the central cluster and surrounding region have an EW in the range 2.5 to 4.5 nm, which translates to an age in the range 5.9 to 6.1 Myr. This is compatible with the findings in \citet[and references therein]{Ohyama1997}, which derives the age from the I(\HeII~$\lambda~4686$)/I(\Hb) vs. EW(\Hb) starburst model diagram. We can also derive the Lyman continuum flux, $N_{Lyc} \approx 1.8 \times 10^{52}$ sec\pwr{-1}  and the number of O7V-type stars, $N_{07V} \approx 3200$ within a radius of 50pc from the peak emission, using the scaling relationships cited in \citet{Brandl2012}. This mixture of old and new stars in the nuclear cluster is by no means unique to this object; this is the case in our own galaxy \citep{Do2009} and in NGC~4244 \citep{Seth2008}.

We can hypothesize that the starburst proceeded outward from the center in a wave, with shock winds from the young stars triggering star formation further out. This would also be supported by the observation that the \Fe{} flux seems to surround the \HII{} region. The ratio of the equivalent width of \HeI{} vs. \brg, for the 150 pc around the nucleus, shows a lower value in the center (1.3--1.5) surrounded by a ring of higher value (1.9--2.1), also indicating a younger stellar population. One could also hypothesize a period of AGN activity providing the initial compression wave.

\subsection{Excitation Mechanisms}
The IR excitation diagram (Fig. \ref{fig:IC630_Excitation}, right-hand panel) shows that almost all pixels are within the ``Starburst'' regime, with minimal identifiable AGN activity (the pixels in the AGN region are located with low absolute flux values, with some corresponding uncertainty). The locations of interest, which are the nucleus, clusters ``1'' and ``3'' and the loci of maximum flux of \Htwo{} and \Fe, are all in the starburst region.

The excitation plot shows a tight correlation for the two line ratios, which is also seen for the nuclear spectrum of active galaxies \citep{Larkin1998, Riffel2013a} as well as on a spaxel-by-spaxel basis \citep{Colina2015}, which concludes that the ISM excitation is determined by the relative flux contribution of the exciting mechanisms and their spatial location. The fit from \citet{Riffel2013a} is log(\Fe/\pab)~=~0.749$\times$log(\Htwo/\brg)-0.207 (plotted in Fig. \ref{fig:IC630_Excitation}), which is consistent, within uncertainties, with the fit from our data (log(\Fe/\pab)~=~0.558$\times$log(\Htwo/\brg)-0.13). That fit also extends over a wider range of excitations, including AGNs and LINERs; their plot also shows that star-forming galaxies are consistently above this fit, as is our fit. The overall nuclear spectrum will have an excitation mode determined by the relative contribution of all sources.

In the optical (BPT) excitation diagram (Fig. \ref{fig:IC630_Excitation_Optical}, right hand panel), again almost all pixels are in the starburst regime. At the NE and SW peripheries, some pixels exhibit ``composite'' excitation, i.e. a mixture of pure starburst and pure AGN (see \citet{Kewley2006} and references therein), indicating there may be some contribution from AGN excitation; this is also possibly present in the infrared along the same axis. We can hypothesize AGN X-ray and shock excitation being obscured towards the line of sight by the starburst ionized outflow, but escaping along the galactic plane. An alternative explanation is presented by \citet{Ohyama1997}, which posits a slowing-down superwind scenario, where the earlier phase of the nuclear starburst generates fast winds ($>$200 \kms) which are now present in the outer regions, as versus the current slower shock velocities in the nucleus as the wind ceases.

Summarizing the excitation for each species:
\begin{itemize}
	\item \HII{} is excited by UV photons from young stars.
	\item \Fe{} is also photo-ionized from the young stars with minor contribution from shocks, as shown by the \Fe/\PII{}~ratios.
	\item \Htwo{}, by contrast, is excited by shocks (from the \Htwo$\lambda~2247.7/\lambda~2121.8$ ratio), with possibly some contribution from X-ray heating from star formation.
\end{itemize}
\clearpage
\subsection{Gas Kinematics}
\subsubsection{Velocity and Channel Maps}
The LOS velocities do not show a simple bipolar structure that would be expected from disk rotation or outflow cones; instead it shows complicated multiple regions of both approaching and receding gas. \citet{Ohyama1997} considers the superwind to be at or nearly face-on to our LOS, which would also explain the lack of any ordered stellar rotation observed (see Fig. \ref{fig:IC630_Stellar_Kinematics}).  The \brg{} and \Fe{} velocity and dispersion fields are virtually identical, both in structure and distribution. The \brg{} velocity distribution (as shown in the velocity histogram) is somewhat bi-modal, peaking around $\pm10$ \kms{}. We suggest that most of the \HI{} gas is in motion and that the apparent velocity at any point is just the mass-weighted sum of the motions towards and away from us. The histograms also show that the \Htwo{} velocity and dipersion is kinematically colder, with the dispersion peak at 30 \kms{}, half that of the \brg{} and \Fe. The \Htwo{} velocity distribution is also much lower.

The \brg{} flux is strongly centrally concentrated, with the LOS velocity presenting a quadrupole pattern and with lower velocity dispersion in the center than at the periphery of the field. The \Fe{} flux is peaked at the periphery of the \brg{} flux; this is compatible with the \Fe{} lines originating in less ionized material than the \HI{} lines. The \Htwo{} is kinematically different; the velocity and dispersion maps and distributions are kinematically cooler.The \Htwo{} velocity map shows almost the same pattern as the \brg{} map, but the N-NW positive velocity region is reduced; this suggests that the two are kinematically de-coupled. This compares with \citet{Riffel2008} for NGC~4051, which also shows a similar channel map; they find the \Htwo{} rotational structure to be dissimilar to the stellar rotation. \citet{Rodriguez-Ardila2004}, in a sample of 22 mostly Seyfert 1 galaxies, suggests that \Fe{} and \Htwo{} originate in different parcels of gas and do not share the same velocity fields, with the \Fe{} flux locations, velocities and dispersions being different to the \Htwo{}; our results support this view. Since IC~630 is close to face-on, it is difficult to determine if the \Htwo{} is in the galactic plane.

The channel maps presents a complex patchy picture (Figs. \ref{fig:IC630_CHMAP_2182}, \ref{fig:IC630_CHMAP_2137} and \ref{fig:IC630_CHMAP_1266}). There is some evidence of structure in the NW (receding) to SE (approaching) directions. Given that the overall gas kinematics (Fig. \ref{fig:IC630_Vel_Sigma}) do not show the signatures of either bi-conal outflow or rotation, we can hypothesize that we are observing the outflows face-on, with streamers of gas rather than the gas filling a complete cone. In this case, the observed structure is again just the mass-weighted sum of the motions. At the extreme velocities, the \brg{} and \Fe{} maps are similar; the difference appears at low velocity, reflecting the over-all flux distribution differences. The \Htwo{} maps are dis-similar; they are kinematically colder. The filamentary structure is similar (though at a smaller scale) to the outflows seen in M82.

\subsubsection{Outflows}
	
Let us examine the geometry of the \HII{} outflow. The half-light radius (i.e. the radius of a circle from the center that covers half the total flux in each channel) increases from 100 pc to 115 pc over the velocity range 0 to $\pm$270 \kms.  Within uncertainties, we can model this as a cylinder, as the half-light radius expands with height, rather than shrinks as a spherical shell expansion would display. The centroid position does not move much, except at extreme velocities where a patchy structure will have greatest effect, showing that the outflow is nearly face-on; this will not affect the outflow calculation, as the flux is low at these velocities.

Calculating the flux times the velocity at each channel gives us a ``momentum'' measure; this is a maximum at $\pm$ 90 \kms{} (equivalent to $9 \times 10^{-5}$ pc yr\pwr{-1} or 90 pc Myr\pwr{-1}) and will be used to calculate the outflow.  The radius of 90\% of the flux for this channel is 205 pc which is an area of $1.3 \times 10^{5}~\text{pc}^{2}$, the flux total in both channels is 10.4 \ecsf{} and the channel width is 60 \kms, which is equivalent to 360 pc. Simplifying the geometry to a cylinder, this is a total volume of $9.5 \times 10^{7}~\text{pc}^{3}$ for both channels. We can calculate the mass of \HII{} in the channel from equation \ref{eqn:HII} above; this is $3.4 \times 10^{5}$ \msun; the density is thus $3.6 \times 10^{-3}$ \msun pc\pwr{-3}. From these values, we obtain $\dot{M}~=~0.09$~\msun/yr. This is in line with the studies of other outflows from Seyferts and LINERs (0.01 -- 10 \msun/yr) \citep{Riffel2015}.

The maximum kinetic power of the outflow is at the +150 \kms{} ($1.8 \times 10^{38}$ erg s\pwr{-1}) and -210 \kms{} ($3.9 \times 10^{38}$ erg s\pwr{-1}) channels. Integrating over all velocity channels, the total power is $1.8 \times 10^{39}$ erg s\pwr{-1}. This is two orders of magnitude below the estimates for Mrk 1157 \citep{Riffel2011} and NGC~4151 \citep{Storchi-Bergmann2010} of 2.3 and 2.4 $\times 10^{41}$ erg s\pwr{-1}, respectively. These galaxies have significant AGN activity, rather than just star formation, to generate larger outflows at higher velocities.
\subsection{Simulations}
\label{sec:SimulCompare}
Schartmann et al.  (2017, in preparation) present simulations of the evolution of circum-nuclear disks in galactic nuclei. 3D AMR hydrodynamical simulations with the RAMSES code \citep{Teyssier2001} are used to self-consistently trace the evolution from a quasi-stable gas disk. This disk undergoes gravitational (Toomre) instability, forming clumps and stars. The disk is subsequently partially dispersed via stellar feedback. The model includes a 10\pwr{7}~\msun{} SMBH, with a $8.5 \times 10^{9}$~\msun{} galactic bulge, and includes SN feedback, which drives both a low gas density outflow as well as a high density fountain-like flow. It finds that the gas forms a three component structure: an inhomogeneous, high density, molecular cold disc, surrounded by a geometrically thick distribution of molecular clouds above and below the disk mid-plane and tenuous, hot, ionized outflows perpendicular to the disk plane. 

\begin{figure}
	\centering
	\includegraphics[width=1.1\linewidth]{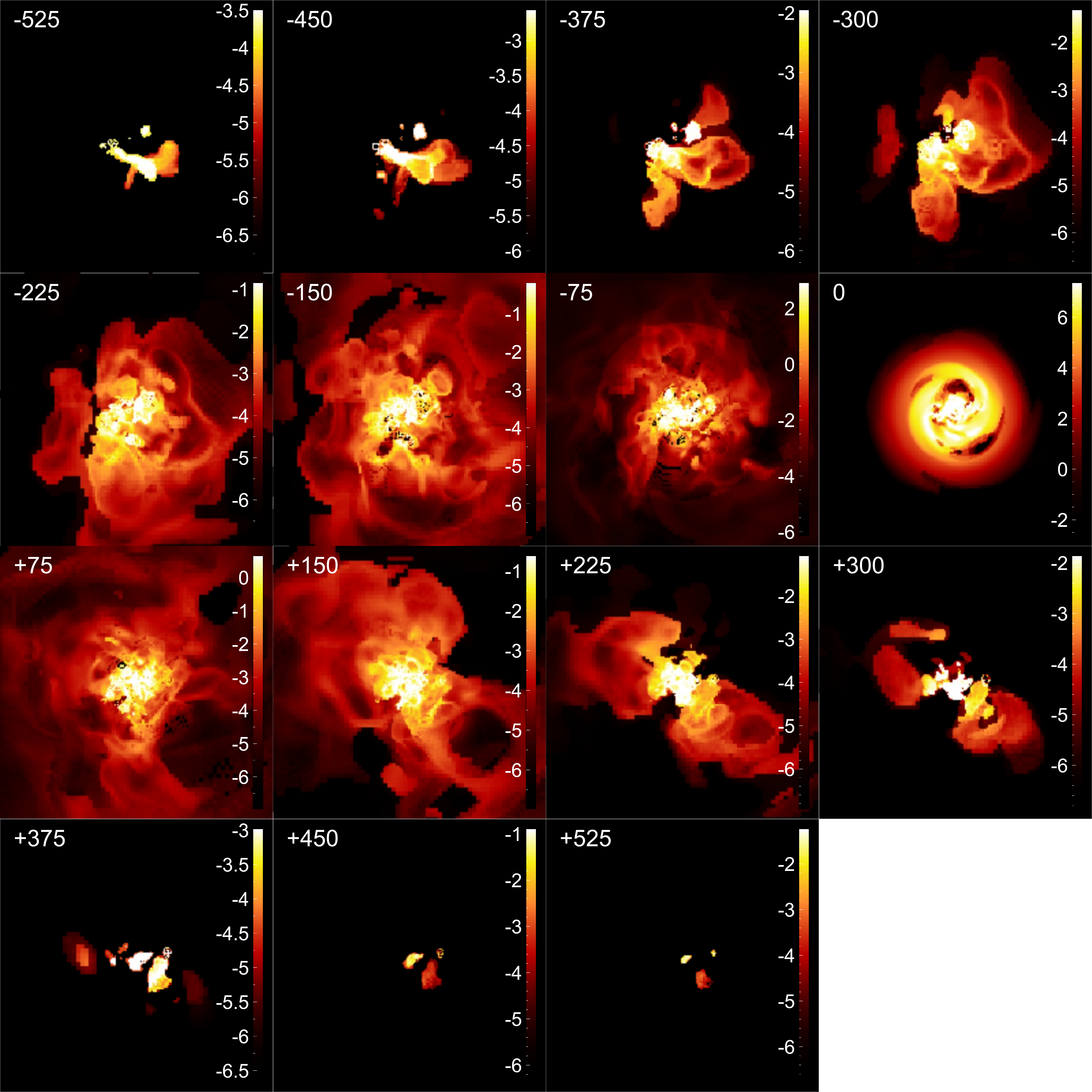}
	\caption{Simulation emissivity channel maps at 30 Myr, in velocity slices of 150 \kms. The values are plotted in log scale of gas number density squared integrated along the line of sight (units of pc cm\pwr{-6}). Maps are 1000$\times$1000 pc; scale bar is shown in the in first map.}
	\label{fig:msgasemissivitychmap}
\end{figure}

Star formation continues until the disk returns to stability. This process cycles over $\sim$200 Myr (depending on initial parameters); the starburst only consumes about half the gas and further inflows will feed the central region until instability is reached and the process can start again. This scenario can explain the observations of short-duration, intense, clumpy starbursts in Seyfert galaxies in their recent ($\sim$10 Myr) history \citep{Davies2007b}. The results from Schartmann et al. show strong similarities to the IC~630 observations, with the initial formation of a few star clusters and clumps, and the outflows in filaments with a broad-based cone or cylindrical geometry reaching to similar scale heights of several hundred pc. The channel map at 30 Myr (Fig. \ref{fig:msgasemissivitychmap}) shows the filamentary structure as seen with IC~630, with coherent organization being traced through the velocity cuts.

\citet{Seth2008} proposed two possible nuclear cluster formation mechanisms: (1) episodic accretion of gas from the disk directly onto the nuclear star cluster; or (2) episodic accretion of young star clusters formed in the central part of the galaxy due to dynamical friction. The simulation results suggest the second scenario as more likely; however the simulation indicates that the clusters are destroyed and redistributed through relaxation to the global potential, rather than dynamical friction.

We can suggest a ``life-cycle'' of nuclear gas, star formation and AGN activity.
\begin{itemize}
	\item Gas flows in to the nucleus of a galaxy, through minor mergers or tidal torques, e.g. bars, where it collects in a disk in the bulge (and/or SMBH) gravitational potential.
	\item The gas disk becomes Toomre-unstable and starts collapsing into star-forming clumps and clusters. The SF rate peaks at about 6--30 Myr after the onset of instability. AGN activity may contribute to the instability.
	\item Stellar feedback, including supernovae and hot OB/WR stellar winds, partially disperses the gas disk and drives filamentary outflows with a scale height of several hundred pc, with the maximum flow at about 10--30 Myr. These winds do NOT fuel any significant AGN activity \citep{Davies2007a}.
	\item At about 150 Myr, star formation declines, with the gas-disk approaching Toomre-stability within the following 20--30 Myr. The stars settle into a nuclear disk about 40--100 pc across.
	\item AGB winds efficiently feed any SMBH and AGN activity starts about 50--100 Myr after the starburst. The SMBH grows by this feeding and tidal friction infall from the gas disk. AGN outflows may trigger further star formation and the activity continues until all available gas is consumed.
\end{itemize}

We have caught IC~630 in the phase between the starburst and AGN activity.

\section{Conclusions}
We have mapped the gas and stellar flux distribution, excitation and kinematics from the inner $\sim$ 300 pc radius of the starburst S0 galaxy IC~630 using NIR J, H and K-band integral-field spectroscopy at a spatial resolution of 37--43 pc ($0.23-0.26\arcsec$), plus additional optical IFS. The main conclusions of this work are as follows.
\begin{itemize}
	\item The nuclear region has a central cluster or disk (half-light radius of 50 pc) with at least two other light concentrations (clusters) within 130 pc. The central stellar population is a mixture of young ($\sim$ 6 Myr) and older stars. 
	\item The stellar kinematics show a SMBH of $M_\bullet = 2.25 (-1.6, +5.1) \times 10^{5} \mmsun{}$ (within 1.4$\sigma$ of there being no central black hole). The AGN-like bolometric luminosity of the galaxy (radio and X-rays) is  mostly from star formation, rather than BH activity.
	\item Within 200 pc of the nucleus the mass of the cold ISM, as derived from extinction, is $M_{ISM} \approx 6.1 \times 10^{6}~\mmsun$, the estimated cold molecular gas mass is $M_{H_{2}(cold)} \approx 3.3 \times 10^{7}~\mmsun$, the mass of ionized gas is $M_{HII} \approx 3.4 \times 10^{6}~\mmsun$ while the mass of the hot molecular gas is $M_{H_{2}} \approx 46~\mmsun$. The cold \Htwo{} mass estimate is greater than that of the ISM as derived from extinction; star formation photo-ionization and the high excitation temperature may have sublimated the dust grains.
	\item The star formation rate is 0.8 \msun/yr, with the SN rate of 1 per 135 years in the central 200 pc, producing X-ray and radio emission and releasing iron from dust grains, which is subsequently photo-ionized and shock heated.
	\item Emission-line diagnostics show that the vast majority of gas excitation is due to star formation, with minimal input from AGN activity. For the main species, \HII{} is excited by UV photons from young stars, \Fe{} is also photo-ionized from the young stars with minor contribution from shocks, whereas \Htwo{} is excited by mainly shocks, with possibly some contribution from X-ray heating from star formation. 
	\item The \Fe{} and \HI{} emissions are closely coupled in velocity and dispersion, but the peak flux of the \Fe{} is at the periphery of the \brg{} flux. The \Htwo{} is kinematically colder than the \HI, and is also spatially located differently to both the \Fe{} and \HI.
	\item A starburst $\sim$ 6 Myr ago provided powerful outflow winds, with the ionized gas outflow rate at 0.09 \msun/yr in a face-on truncated cone geometry. 
	\item Our observations are broadly comparable with simulations where a Toomre-unstable gas disk triggers a burst of star formation, peaking after about 30 Myr and possibly cycling with a period of about 200 Myr.
\end{itemize}
IC~630 is an example of a galaxy that has ``AGN-like'' activity (radio and X-ray emission) but displays minimal AGN excitation. Even though it is an S0 galaxy, it has a high star formation rate in both the central region and over the whole galaxy. This object is an example of nuclear star-formation dominating the narrow-line region emission. The nuclear young stars and SN providing photo-ionization, stellar winds and shocks to excite the \HI{}, \Htwo{} and \Fe{}.



\acknowledgments
This study is based on observations obtained at the Gemini Observatory, which is operated by the Association of Universities for Research in Astronomy, Inc., under a cooperative agreement with the NSF on behalf of the Gemini partnership: the National Science Foundation (United States), the National Research Council (Canada), CONICYT (Chile), Ministerio de Ciencia, Tecnolog\'{i}a e Innovaci\'{o}n Productiva (Argentina), and Minist\'{e}rio da Ci\^{e}ncia, Tecnologia e Inova\c{c}\~{a}o (Brazil).  It is also based on observations made with ESO Telescopes at the Paranal Observatory. The respective program IDs are given in section \ref{sec:Observations} above.

This research has made use of the NASA/IPAC Infrared Science Archive, which is operated by the Jet Propulsion Laboratory, California Institute of Technology, under contract with the National Aeronautics and Space Administration. The SIMBAD Astronomical Database is operated by CDS, Strasbourg, France. IRAF is distributed by the National Optical Astronomy Observatories,which is operated by the Association of Universities for Research in Astronomy, Inc. (AURA) under cooperative agreement with the National Science Foundation. The National Radio Astronomy Observatory is a facility of the National Science Foundation operated under cooperative agreement by Associated Universities, Inc.

The Pan-STARRS1 Surveys (PS1) and the PS1 public science archive have been made possible through contributions by the Institute for Astronomy, the University of Hawaii, the Pan-STARRS Project Office, the Max-Planck Society and its participating institutes, the Max Planck Institute for Astronomy, Heidelberg and the Max Planck Institute for Extraterrestrial Physics, Garching, The Johns Hopkins University, Durham University, the University of Edinburgh, the Queen's University Belfast, the Harvard-Smithsonian Center for Astrophysics, the Las Cumbres Observatory Global Telescope Network Incorporated, the National Central University of Taiwan, the Space Telescope Science Institute, the National Aeronautics and Space Administration under Grant No. NNX08AR22G issued through the Planetary Science Division of the NASA Science Mission Directorate, the National Science Foundation Grant No. AST-1238877, the University of Maryland, Eotvos Lorand University (ELTE), the Los Alamos National Laboratory, and the Gordon and Betty Moore Foundation.

The national facility capability for SkyMapper has been funded through ARC LIEF grant LE130100104 from the Australian Research Council, awarded to the University of Sydney, the Australian National University, Swinburne University of Technology, the University of Queensland, the University of Western Australia, the University of Melbourne, Curtin University of Technology, Monash University and the Australian Astronomical Observatory. SkyMapper is owned and operated by The Australian National University's Research School of Astronomy and Astrophysics. The survey data were processed and provided by the SkyMapper Team at ANU. The SkyMapper node of the All-Sky Virtual Observatory is hosted at the National Computational Infrastructure (NCI).

We would like to thank our Triplespec colleagues at Caltech for their ongoing support of our survey. We would also like to thank the anonymous reviewer for a very detailed and constructive report.

JM, MD and MS acknowledge the continued support of the Australian Research Council (ARC) through Discovery project DP140100435.

\facilities{Gemini:Gillett (NIFS), VLT:Yepun (SINFONI), ATT (WiFeS)}
\bibliographystyle{aasjournal}
\bibliography{IC630_ArXiv.bbl}
\end{document}